\documentclass[fleqn,usenatbib]{mnras}
\usepackage{newtxtext,newtxmath}

\usepackage[T1]{fontenc}

\DeclareRobustCommand{\VAN}[3]{#2}
\let\VANthebibliography\thebibliography
\def\thebibliography{\DeclareRobustCommand{\VAN}[3]{##3}\VANthebibliography}

\usepackage{graphicx}	
\usepackage{amsmath}	

\usepackage{amssymb}	

\usepackage{booktabs}
\usepackage{multirow}
\usepackage[normalem]{ulem}
\newcommand{\me}{  {$\mathcal{M}_{e}$}}
\newcommand{\Mnom} {{M$_{\odot}$}}

\title[BSSs in open clusters using Gaia]{Blue Straggler Stars in Open Clusters using \textit{Gaia}: Dependence on Cluster Parameters and Possible Formation Pathways}

\author[Jadhav and Subramaniam]{
Vikrant V. Jadhav,$^{1,2}$\thanks{E-mail: vikrant.jadhav@iiap.res.in}
and Annapurni Subramaniam,$^{1}$
\\
$^{1}$Indian Institute of Astrophysics, Koramangala II Block, Bangalore-560034, India\\
$^{2}$Joint Astronomy Program and Physics Department, Indian Institute of Science, Bangalore-560012, India
}

\date{Accepted XXX. Received YYY; in original form ZZZ}

\pubyear{2021}

\begin{document}
\label{firstpage}
\pagerange{\pageref{firstpage}--\pageref{lastpage}}
\maketitle

\begin{abstract}
Blue straggler stars (BSSs) are the most massive stars in a cluster formed via binary or higher-order stellar interactions. Though the exact nature of such formation scenarios is difficult to pin down, we provide observational constraints on the different possible mechanism. In this quest, we first produce a catalogue of BSSs using \textit{Gaia} DR2 data. Among the 670 clusters older than 300 Myr, we identified 868 BSSs in 228 clusters and 500 BSS candidates in 208 clusters. In general, all clusters older than 1 Gyr and massive than 1000 \Mnom\ have BSSs. The average number of BSSs increases with cluster age and mass, and there is a power-law relation between the cluster mass and the maximum number of BSSs in the cluster. We introduce the term fractional mass excess (\me) for BSSs. We find that at least 54\% of BSSs have \me\ $<$ 0.5 (likely to have gained mass through a binary mass transfer (MT)), 30\% in the $1.0 <$ \me\ $< 0.5$ range (likely to have gained mass through a merger) and up to 16\% with \me\ $>$ 1.0 (likely from multiple mergers/MT). We also find that the percentage of low \me\ BSSs increases with age, beyond 1--2 Gyr, suggesting an increase in formation through MT in older clusters. The BSSs are radially segregated, and the extent of segregation depends on the dynamical relaxation of the cluster. The statistics and trends presented here are expected to constrain the BSS formation models in open clusters.
\end{abstract}



\section{Introduction} \label{sec:Introduction}

Blue straggler stars (BSSs) are one of the most massive stars in a cluster. They stand out from other cluster members due to their bluer and brighter position in the colour-magnitude diagram (CMD). Since their first reporting \citep{Sandage_1953AJ.....58...61S}, multiple mechanisms have been proposed for their formation. As all the stars in a cluster formed at almost the same time, there should not be stars above the main sequence (MS). This apparently longer life is justified by some type of mass accretion by the progenitor of the BSS. The major formation pathways are as follows:
\begin{enumerate}
    \item \citet{McCrea_1964MNRAS.128..147M} proposed that mass transfer (MT) from a binary companion can lead to rejuvenation of the acceptor and formation of a BSS. The MT efficiency depends on the orbital periods: wider orbits have non-conservative MT and leave a remnant behind, while close binaries can have conservative MT and lead to mergers.
    Depending on the type of MT (case-A/B/C), this results in BSS+white dwarf (WD) (lower mass to normal mass WD). Signatures of hotter/compact companion can detect such systems (deconvolving spectral energy distributions (e.g. \citealt{Sindhu2019ApJ...882...43S}) and variability in radial velocity).
    MT systems are typically expected to have circular orbits due to the past MT event, but MT in elliptical orbits has also been seen (e.g. \citealt{Boffin2014A&A...564A...1B}). 
    \item Collisions of individual stars or mergers from collisions of binary pairs are also linked to the BSS formation \citep{Hills_1976ApL....17...87H,Leonard_1989AJ.....98..217L}. Collisions typically happen in dense environments (such as globular clusters) and do not show chemical peculiarities of an MT event \citep{Sills_1999ApJ...513..428S}. However, \citet{Leigh_2013MNRAS.428..897L} found no correlation between collision rate and number of BSSs, meaning collisions are not the dominant pathway, even in dense globular clusters.
    \item \citet{Naoz_2014ApJ...793..137N} showed that the eccentric Kozai--Lidov mechanism could tighten the inner binary in a hierarchical triple system. \citet{Perets_2009ApJ...697.1048P} suggested that such merger of inner binary has a significant role in BSSs formation in open clusters (OCs). Presently, such systems will likely be MS+BSS binaries with long eccentric orbits. 
\end{enumerate}

However, different mechanisms are said to dominate in different cluster environments: 
(i) less dense clusters favour binary MT pathway while high-density clusters favour collisional pathway \citep{Davies_2004MNRAS.349..129D}, (ii) binary pathway is dominant in globular clusters of all masses  \citep{Leigh_2007ApJ...661..210L, Knigge_2009Natur.457..288K}, (iii) core collapse of a globular cluster can trigger a burst of BSS formation \citep{Ferraro_2009Natur.462.1028F}, (iv) old, less dense and relaxed clusters favour binary pathway \citep{Mathieu_2015ASSL..413...29M}.

As mentioned above, there are different mechanisms responsible for forming BSSs, and it is not easy to pinpoint the exact pathway without information about their companions, chemical signatures and orbital parameters. 
There are many dedicated studies on the BSS population of individual clusters, but only a few on BSS population across different clusters (e.g. \citealt{Leiner_2021ApJ...908..229L}). 
To study the BSSs in a larger cluster sample, we require a homogeneous catalogue. \citet{Ahumada_2007A&A...463..789A} had provided the pre-\textit{Gaia} catalogue of BSSs and recently \citet{Rain_2021arXiv210306004R} have produced a \textit{Gaia} based BS catalogue (though it is not available at the time of this writing).
A homogeneous catalogue of BSSs will help analyse the dependence of the BS population on the cluster properties and occurrence in them. Furthermore, such a catalogue can be further used to make targeted observations of exotic BS population in the multi-wavelength regime, revealing and constraining the formation scenarios and stellar parameters.

Multi-wavelength studies of BSSs using Hubble Space Telescope \citep{Gosnell2014ApJ...783L...8G, Gosnell2019ApJ...885...45G} and the Ultra-violet Imaging Telescope (UVIT) on \textit{AstroSat} \citep{Subramaniam2016ApJ...833L..27S, Sindhu2019ApJ...882...43S, Jadhav2021arXiv210213375J} have detected hot companions to BSSs. The hot companions of the BSSs were found to differ from cluster to cluster and found to show a large variety, such as extremely low-mass WD, horizontal branch (HB) stars and sub-dwarfs. These studies suggest that remnants of the donors in binary BSSs occupy a fairly large parameter space that is yet to be explored properly. One of the aims of this study is to identify potential clusters with BSSs to be followed up with UV imaging using \textit{AstroSat}. 

In this study, we have created a catalogue of BSSs in OCs (\S \ref{sec:Data_and_Method}), presented and discussed the properties of BSSs and their dependence on cluster parameters (\S \ref{sec:Results} and \S \ref{sec:discussion}).

\section{Data and Classification} \label{sec:Data_and_Method}

As BSSs lie above the MS turnoff (MSTO) of the clusters, where stars are normally not expected to be found, accurate proper motion information is necessary to confirm their membership. \citet{Cantat2018,Cantat2020} provided membership and cluster parameters of OCs using \textit{Gaia} DR2 \citep{Gaia2016A&A...595A...1G, Gaia2018A&A...616A...1G}. Their catalogue provides the age, distance, extinction and other aggregate cluster properties. We selected 670 clusters with log($age$) over 8.5 ($\sim$300 Myr) as our primary sample. The lower age cutoff was chosen to avoid confusion between BSSs and blue giants in younger clusters. Though it is known that the BSSs like systems are more prevalent in relatively older clusters, it is also required to explore their existence in younger clusters. The clusters in the age range 300 Myr--10 Gyr will serve both purposes.

\begin{figure}
    \centering
    \includegraphics[width=0.48\textwidth]{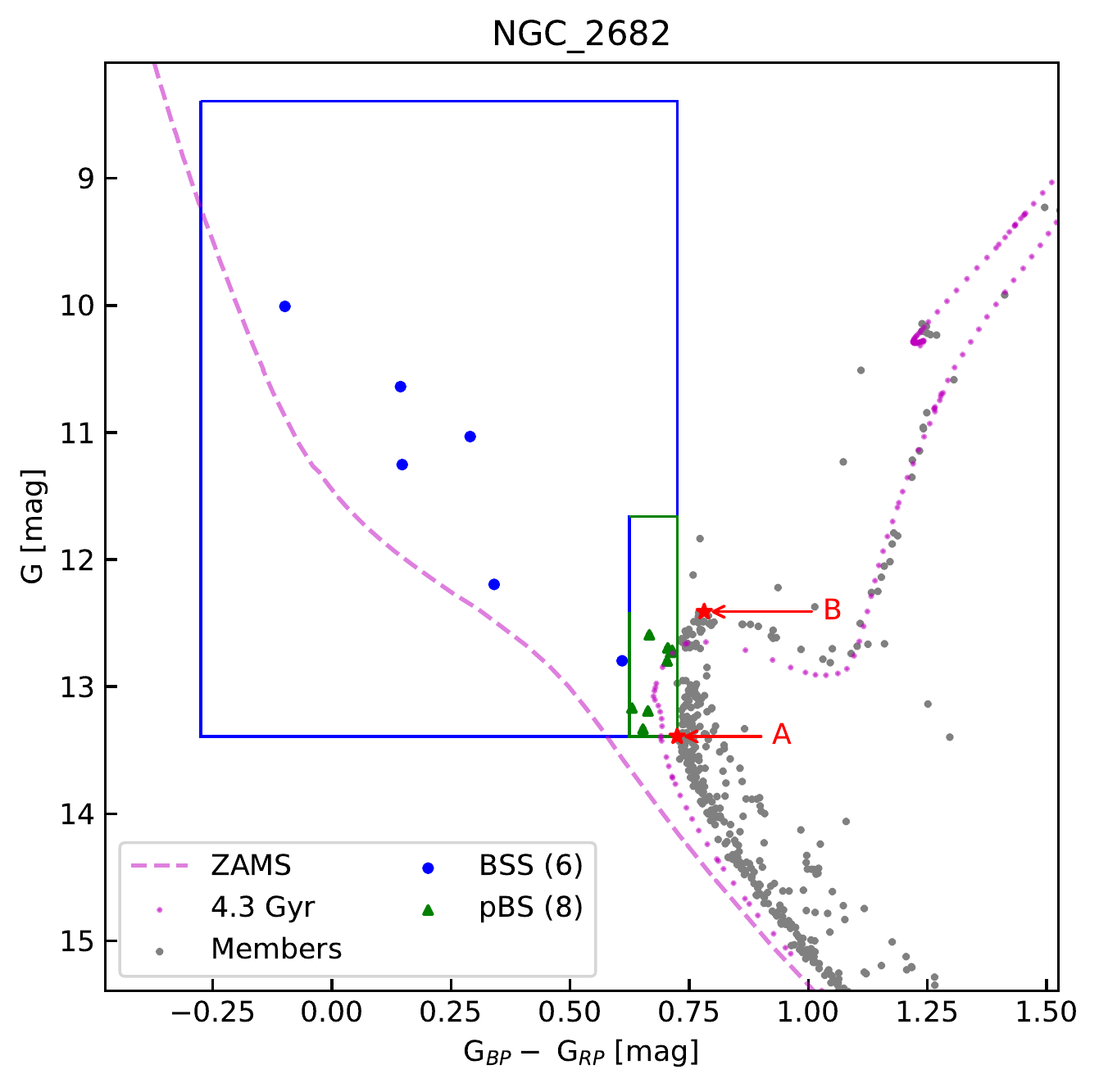}
    \caption{Schematic of classifying BSSs and pBSs in the colour-magnitude plane. Stars in blue and green boxes are classified as BSS and pBS respectively. The isochrone and ZAMS are shows for comparison. The manually identified A and B points are shown as red stars.}
    \label{fig:demo_class}
\end{figure}

Detection of BSSs requires a clear identification of the cluster MSTO. Although \citet{Cantat2020} listed the cluster parameters, identifying accurate MSTO is not trivial due to differences between isochrone models and the neural network used by \citet{Cantat2020}. The same can be seen in Fig.~\ref{fig:demo_class}, where the cluster data and the solar metallicity PARSEC\footnote{http://stev.oapd.inaf.it/cgi-bin/cmd} isochrone \citep{Bressan2012MNRAS.427..127B} differ by $\sim$ 0.1 mag in the colour axis near the MSTO. Fig.~\ref{fig:example_CMDs} also shows that isochrone fits of NGC 2420 and 2506 are not good enough to locate the turnoff point, while isochrone MSTO of NGC 2243 and 6791 are quite similar to the actual data. Another approach with an automated search of the bluest MS point in the CMD works well in rich clusters, but not particularly well in poorer clusters. Hence, we have manually selected the MSTO point (A; bluest point on the cluster CMD)
and the brightest point in the MS (B; brightest point on the CMD, before starting of the sub-giant branch).

\begin{figure}
    \centering
    \includegraphics[width=0.48\textwidth]{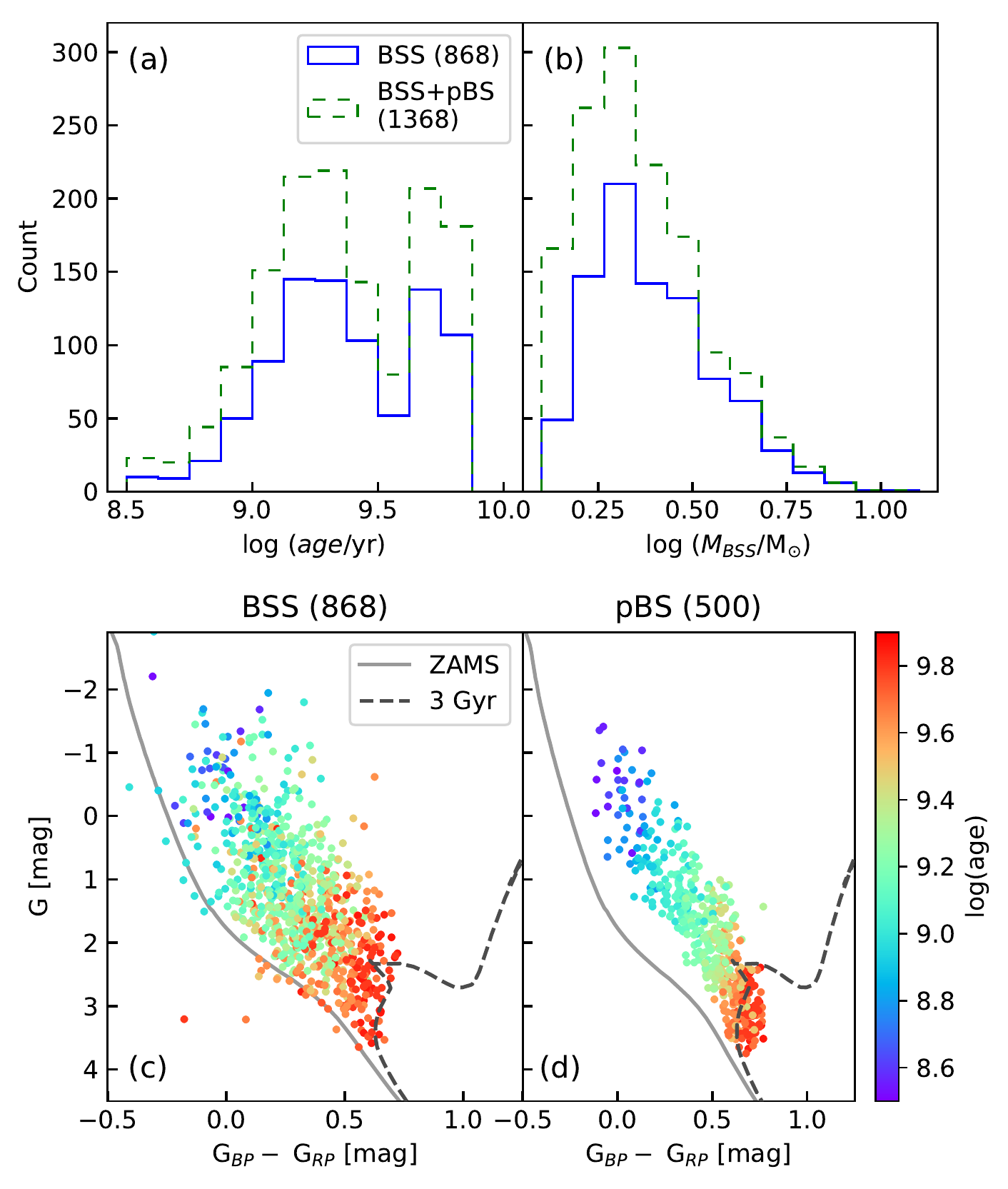}
    \caption{(a) Age and (b) mass distribution of the BSSs as blue histograms (and BSS+pBS as dashed green histograms). CMDs of (c) BSSs and (d) pBSs coloured according to the cluster age. All stars are corrected for distance and reddening using cluster parameters in \citet{Cantat2020}. An isochrone of 3 Gyr and ZAMS are shown for reference.}
    \label{fig:BSS_CMDs}
\end{figure}

\begin{figure*}
    \centering
    \includegraphics[width=0.98\textwidth]{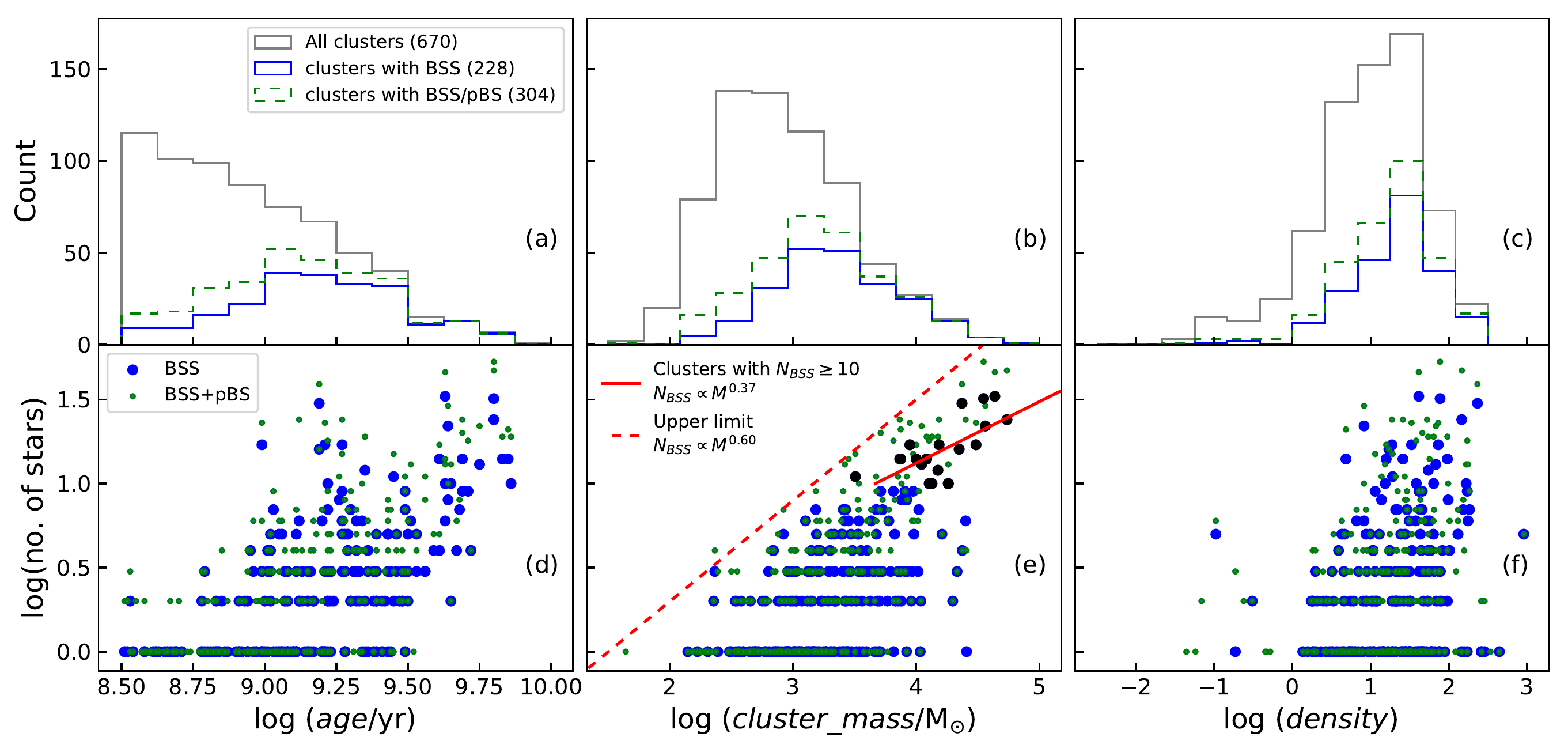}
    \caption{\textit{Upper panels} show the histograms of cluster (a) age, (b) mass and (c) density for all clusters (grey), clusters with BSSs (blue) and clusters with BSS/pBS (dashed green). \textit{Bottom panels} show the dependence of number of BSS (as blue circles) and BSS+pBS (as green dots) on the cluster (d) age, (e) mass and (f) density. The panel (e) shows a linear fit to clusters with at least 10 BSS (black filled circles) as a solid red line. The dashed red line represents the empirical upper limit on the number of BSSs in a cluster based on cluster mass.}
    \label{fig:cl_params}
\end{figure*}

BSSs are expected to be brighter and bluer than the MSTO with a significant range. 
Near the MSTO, there is an inherent scatter in the data, sometimes shifting the MSTO stars to the bluer side. Similarly, some stars in the tip of the MS (e.g. stars with G mag $\sim$ 12.8 in Fig.~\ref{fig:demo_class}) are bluer and brighter than the MSTO. Hence, we created two regions in the CMD to select BSSs and probable BSSs (pBSs) as shown in the blue and green boxes in Fig.~\ref{fig:demo_class}. The blue box (to select BSSs) spans about 1 mag in colour axis and 5 mag in magnitude axis, with respect to point A. 
The green box (to select pBSs) spans 0.1 mag in colour axis and has a bright limit at 0.75 mag above the point B (to avoid any unresolved binaries brighter than the MS tip). In Fig.~\ref{fig:demo_class}, the fainter pBSs in NGC 2682 are genuine BSSs, while brighter pBSs are MSTO stars in the MS tip.  All CMDs are manually checked and the points A and B are identified. The CMDs shown in Fig.~\ref{fig:example_CMDs} demonstrate the need for a  manual selection. It also demonstrates how A and B points are identified uniformly across all the clusters studied here. We believe that this method will greatly help in identifying the BSSs (in the blue box) and the pBSs (in the green box) similarly across the clusters, resulting in a reliable sample. Overall, the BSS class consist of bona fide BSSs, while pBS class consists of some BSSs and some MS stars. We, therefore, mainly use only the BS population for further analysis to study their properties and dependency on cluster parameters.


\section{Results} \label{sec:Results}

Among the 670 clusters older than 300 Myr, we found 868 BSSs in 228 clusters and 500 pBSs in 208 clusters. Overall, 304 clusters have 1368 BSS candidates 
(BSS+pBS).
Table~\ref{tab:cluster_list} shows the first five rows of the table of all 670 clusters along with the cluster properties and number of BSSs and pBSs. Table~\ref{tab:catalogue} shows the example of the list of all BSSs and pBSs. The full tables are available online and in CDS\footnote{\url{http://cdsarc.u-strasbg.fr/viz-bin/qcat?J/MNRAS}}.

\begin{table}
    \centering
    \begin{tabular}{cccc|ccc}
    \toprule
\multicolumn{2}{c}{Binning type}			&	No. of 	&	No. of	&	\multicolumn{3}{c}{No. of BSSs in \me\ classes}					\\	\cmidrule(lr){5-7}
	&		&	Clusters	&	BSSs	&	low	&	high	&	extreme	\\	\hline
\multirow{5}{*}{\rotatebox[origin=c]{90}{\parbox[c]{2cm}{\hfill log($age$)}}}	&	8.50--9.00	&	56	&	90	&	43	&	27	&	20	\\	
	&	9.00$-$9.25	&	77	&	234	&	109	&	71	&	54	\\	
	&	9.25$-$9.50	&	65	&	247	&	121	&	84	&	42	\\	
	&	9.50$-$9.75	&	24	&	190	&	123	&	49	&	18	\\	
	&	9.75$-$10.0	&	6	&	107	&	75	&	29	&	3	\\	\hline
\multirow{5}{*}{\rotatebox[origin=c]{90}{\parbox[c]{2cm}{\hfill log($cl\_mass$)}}}	&	2.0$-$3.0	&	55	&	85	&	44	&	28	&	13	\\	
	&	3.0$-$3.5	&	88	&	207	&	112	&	61	&	34	\\	
	&	3.5$-$4.0	&	60	&	272	&	140	&	83	&	49	\\	
	&	4.0$-$4.5	&	21	&	193	&	101	&	56	&	36	\\	
	&	4.5$-$5.0	&	4	&	111	&	74	&	32	&	5	\\	\hline
	&	Total	&	228	&	868	&	471	&	260	&	137	\\
	\bottomrule
    \end{tabular}
    \caption{The distribution of BSSs in different \me\ classes for clusters binned by age and mass. The first column shows the limits of cluster age and mass for each bin. Second and third columns have the number of clusters and BSSs in the respective bins. Last three columns show the number of BSSs divided into low-\me, high-\me\ and extreme-\me\ BSSs. The last row shows the \me\ class distribution of all BSSs.}
    \label{tab:binwise}
\end{table}

The age-wise distribution of 228 clusters with 868 BSSs is given in Table~\ref{tab:binwise} and is shown in Fig.~\ref{fig:BSS_CMDs} (a). {Among the given bins, the largest number of clusters (77) in this study lies in the log(\textit{age}) 9.00--9.25 bin, whereas the largest number of BSSs (247) lies in the 9.25--9.50 bin.} 
Fig.~\ref{fig:BSS_CMDs} (c) \& (d) shows the absolute CMDs of selected BSSs and pBSs, coloured according to the cluster age. The majority of the BSSs are found to be located redder than the zero-age MS (ZAMS).
The pBSs are found as a neat sequence, mainly due to the selection criteria used. Appendix \ref{sec:appendix_A} presents the details of estimation of the following cluster parameters: luminosity function (LF), cluster mass, effective radii, relaxation time and cluster density, along with the mass estimation of BSSs. We discuss the properties of the BSSs, pBSs and the cluster parameters below.

\begin{figure*}
    \centering
    \includegraphics[width=0.98\textwidth]{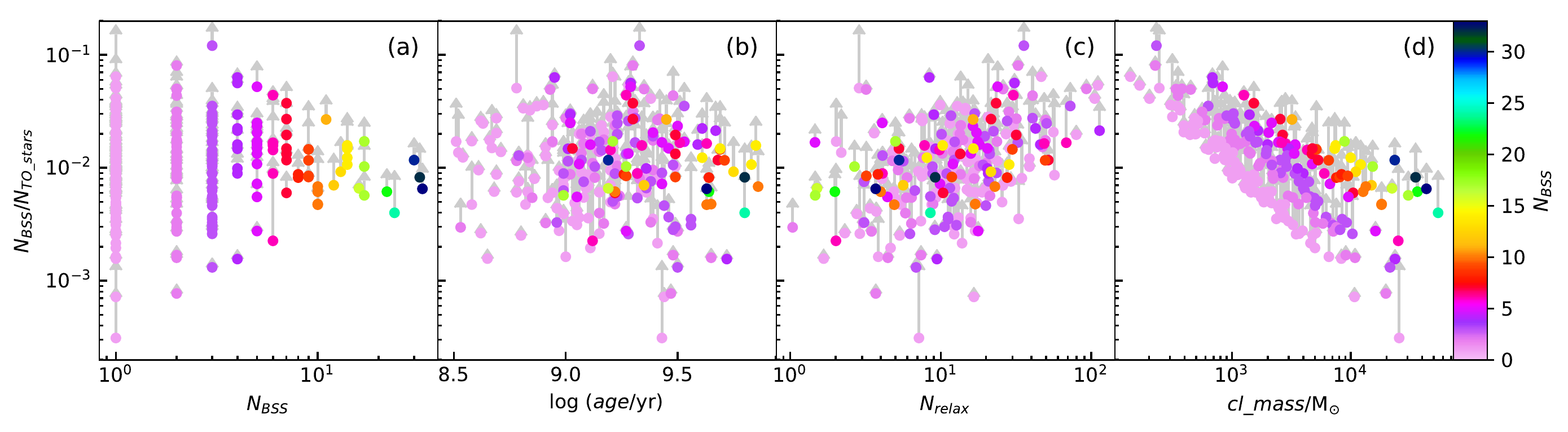}
    \caption{Variation of BSS fraction with number of BSSs, cluster age, relaxation periods passed and mass. The increase in fraction after including pBS is indicated by arrows. The clusters are coloured according to the number of BSSs.}
    \label{fig:fractional_BSS}
\end{figure*}

\subsection{Relation between BSSs and cluster properties}
The age, mass and density distributions of clusters in this study are shown in Fig.~\ref{fig:cl_params} (a)--(c). The figure shows the distribution of all clusters (grey), those with BSSs (blue) and those with BSSs+pBSs (dashed green), as a function of three parameters, to understand their influence. In all the plots, the blue and the green distributions closely resemble each other. 
Figure (a) shows that only a small fraction of clusters younger than 1 Gyr have BSSs, while almost all clusters above 3 Gyr have BSSs. We notice an increasing trend in the fraction of clusters to have BSSs, in the range 300 Myr to 3 Gyr. We also notice that the number of clusters with BSSs (and pBSs) increases up to 1 Gyr, shows a flat peak in the 1--3 Gyr age range among the sample studied here. 

In this sample studied here, we have most of the clusters in the log(\textit{mass}) range 3.0--3.5, whereas the majority of the BSSs are found in the log(\textit{mass}) range of 3.5--4.5. Only a small fraction of clusters lighter than 1,000 \Mnom\ have BSSs, while almost all clusters massive than 10,000 \Mnom\ have. Clusters with mass in the range 100--10,000 \Mnom\ show an increasing fraction. The number of clusters with BSSs (and pBSs) increases up to 1,000 \Mnom and has a flat peak between 1000--3000 \Mnom.  Any such definite trend is not reflected in the density plot, where clusters with a wide range in density show the presence and absence of BSSs in the clusters. The distribution of clusters with BSSs seems to be a subset of the overall sample distribution, where denser clusters are likely to have more BSSs. 

Fig.~\ref{fig:cl_params} (d)--(f) shows the relation of the cluster age, mass and density with the number of BSSs (blue) and BSSs+pBSs (green).
Here we have shown all clusters (with BSSs and pBSs) such that each blue dot denotes the number of BSSs in the cluster and green dot denotes the total number (BS+pBSs) against the cluster parameter, where the green dots can be considered as the upper limit.  
As expected, the older clusters, massive clusters and more dense clusters have more BSSs. This is apparent from the rising trend seen in panels (d) to (f). 
If we look at the table~\ref{tab:binwise}, we find an interesting doubling trend in the BSS frequency with age. We find that, on an average, clusters in log(\textit{age}) = 8.5--9.0 range, have 1.6 BSS/cluster. This doubles to 3.0 BSS/cluster in the range 9.0--9.25, which stays more or less similar (3.8 BSS/cluster) in the 9.25--9.5 range. This again doubles to 7.9 BSS/cluster in the 9.5--9.75 range and again doubles (and maybe a bit more, 17.8 BSS/cluster) in the 9.75--10.0 range. For the entire sample, we get 3.8 BSS/cluster across all age range.

Interestingly, the dashed line in Fig.~\ref{fig:cl_params} (e) shows that there is an upper limit on the number of BSSs present in a cluster depending on the cluster mass, as follows: 
\begin{equation}
\label{eq:NBSS_mass}
    \textrm{log} (N_{BSS,max}) = 0.6\,\textrm{log}(cl\_mass/M_{\odot}) - 0.9
\end{equation}
This relation is estimated incorporating both the BSSs and the pBSs. This relation seems to work well for clusters more massive than 500 \Mnom. From the table~\ref{tab:binwise}, we find a large range in the number of BSS per cluster from 1.5 BSS/cluster in the log(\textit{mass}) range 2.0--3.0, which increases to 2.6 (3.0--3.5 range), 4.6 (3.5--4.0), 9.25 (4.0--4.5), reaching finally to 26.75 BSS/cluster (4.5--5.0). The upper limit is much higher than that for the older clusters (i.e., 17.8). 

In panel (f), though we do detect a tentative rising trend with respect to density, such that the more dense clusters produce more BSSs, the observed scatter stops us from deriving a relation.
To compare the BSS frequency across different clusters, we calculated BSS fraction by normalising with MSTO population ($N_{BSS}/N_{TO\_stars}$).
Fig.~\ref{fig:fractional_BSS} (a) shows that the fraction has large range in clusters with less BSSs, and it stabilises for clusters with more BSSs with median fraction of 1\%. There is no correlation between the presence of BSSs and the cluster age. 
The clusters with fewer BSSs show linear dependence between BSS fraction and the mass, but that is an artefact due to the range of values available for numerator ($N_{BSS}$) and denominator ($N_{TO\_stars}$). For massive clusters, the BSS fraction tends towards the median value.
Fig.~\ref{fig:fractional_BSS} (c) shows the relaxed clusters tend to have higher BSS fraction.
This could be partly due to evaporation of lower mass stars in relaxed clusters, or more likely due to the inverse relation between cluster mass and relaxation.
Furthermore, we found no correlation between the presence of BSSs and the radius of the cluster, cluster distance, Galactocentric position or number of stars near the turnoff.

\begin{figure*}
    \centering
    \includegraphics[width=0.98\textwidth]{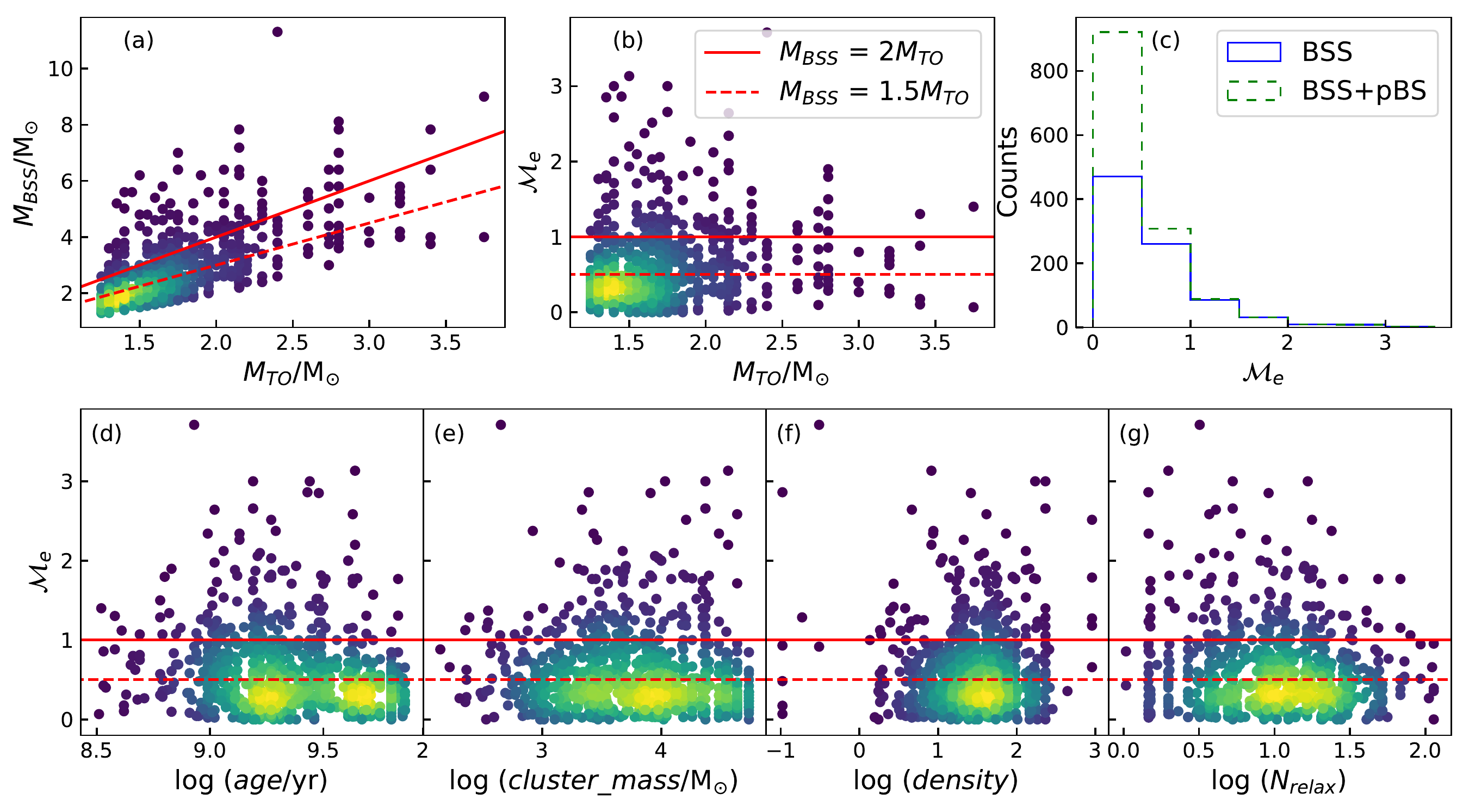}
    \caption{(a) 2-D density plots for visualising the variation of mass of the BSSs with respect to the cluster MSTO mass, where the points are coloured according to the crowding (yellow means crowding). The dashed, and solid red lines represent the BSSs with 1.5$M_{TO}$ and 2$M_{TO}$ mass, respectively. (b) Variation of \me\ with cluster MSTO mass. (c) \me\ distribution of the BSSs and pBSs. \textit{Bottom panels:} Variation of \me\ with (d) cluster age, (e) cluster mass and (f) cluster density and (g) dynamical relaxation.}
    \label{fig:me_dist}
\end{figure*}

\subsection{Fractional mass excess}\label{sec:mass_excess}
The BSSs identified here are substantial in number across a fairly large age range, and therefore the sample has the potential to throw light on their formation mechanisms.
As BSSs have gained mass during their MS lifetime, the excess mass they have gained could reveal their formation pathway. In order to achieve this, we need first to estimate the mass of the identified BSSs and then make an estimate of the mass possibly gained by a BSS. The method used here to estimate the mass of the BSSs using ZAMS is given in Appendix \ref{sec:star_mass}. Fig.~\ref{fig:BSS_CMDs} (b) shows the distribution of mass of the BSSs. Combining the age (a) and the mass (b) plots, we find that the majority of the BSSs are older ($>$ 1 Gyr) and low mass ($<$ 3 \Mnom), but the overall range in age and mass is quite large (0.3--7 Gyr; 1--15 \Mnom). 

The next step then would be to understand how different formation pathways increase the mass of a star. 
In the case of a binary MT, there is a wide range in MT efficiencies: $\leq$ 0.2 in case-B MT, 0.2--0.7 in case-A MT, $\sim$ 1 in conservative MT scenarios \citep{Shao2016ApJ...833..108S}. Typically, wider binaries have non-conservative MT (efficiency $<$ 0.5) and can leave a remnant of the donor. On the other hand, close binaries can have more conservative MT (efficiency $>$ 0.5) and can lead to mergers. 
Although it is not enough to use a single value of MT for describing all binary systems \citep{Mink2007A&A...467.1181D}, we can get an idea about the different pathways by comparing the mass of the BSSs to the mass of the cluster MSTO. As the host clusters have a range in age, they also have a range in the stars that can gain mass and become a BSS. Therefore, it is required to use fractional mass excess, based on a quantity that can be used for normalising, such as the mass at the MSTO (M$_{TO}$). We use the mass of a ZAMS star of the same magnitude as the MSTO (a conservative estimate of the MSTO mass) to determine M$_{TO}$. More discussion on the normalising mass is given in \S \ref{sec:reference_mass}.

Fig.~\ref{fig:me_dist} (a) shows the variation of $M_{BSS}$ with $M_{TO}$. Here, $M_{TO}$ is defined as the mass of the ZAMS star with the same magnitude as the MSTO.
The dashed, and solid red line represents the $M_{BSS} = 1.5M_{TO}$ and $M_{BSS} = 2M_{TO}$, respectively. For example, the solid line can be considered the position of BSSs formed by adding up the mass of two MSTO stars. It is clear from the plot that the majority of the stars follow the dashed line. Also, most of the BSS population is within the solid line. 

The $M_{BSS}$ and $M_{TO}$ have a large range (1--4 \Mnom) in our cluster sample. In order to identify the BSSs that a binary MT could have formed, we need to estimate the fractional mass gain of BSSs, in order to correlate with the efficiency of MT. Hence, we defined a new parameter called `fractional mass excess' (\me) which normalises the $M_{BSS}$ with the cluster $M_{TO}$:
\begin{equation}
    \mathcal{M}_{e} = \frac{M_{BSS}-M_{TO}}{M_{TO}}
\end{equation}
By definition, \me\ is equivalent to the MT efficiency in the case where both accretor and progenitor were MSTO stars. 
Fig.~\ref{fig:me_dist} (b) shows the relation between \me\ and  $M_{TO}$ and \me\ can directly give information about the MT efficiency in clusters with any $M_{TO}$. The plot shows that typically $M_{TO}$ is less than 2.0 \Mnom. Moreover, most of the BSSs are below \me\ $<$ 0.5.
Fig.~\ref{fig:me_dist} (c) shows the histogram of \me\ for all BSSs and pBSs in our sample. The peak of the distribution is in the 0--0.5 range, with most of the BSSs having \me\ $<$ 1, and a few having large \me\ in the range of 1--3.5. We also note that the large \me\ values are mostly present in clusters with $M_{TO}$ $<$ 2.5 \Mnom. 

Hereafter we classify BSSs as low-\me\, high-\me\ and extreme-\me\ BSSs for \me\ $<$ 0.5,  0.5 $<$ \me\ $<$ 1.0 and \me\ $>$ 1.0 respectively. Overall, there are 471 (BSSs)--921 (BSS+pBS) (54--67\%) low-\me\ BSSs, 260--308 (30--22\%) high-\me\ BSSs and 137--139 (16--10\%) extreme-\me\ BSSs. As expected, most of the BSSs fall within the low to high-\me. However, we find that there is a non-negligible fraction of extreme-\me\ BSSs, that requires attention.

Fig.~\ref{fig:me_dist} (d)--(g) present the variation of \me\ with the cluster age, mass, density and number of relaxation periods passed. We find the low-\me\ BSSs to be the majority in all clusters. The very young and very old clusters appear to have less extreme-\me\ BSSs. In the (d) panel, two yellow blobs identify two peaks in the BSSs (similar to the distribution in Fig.~\ref{fig:demo_class} (a)), though likely to be a selection effect, suggests that the low-\me\ BSSs dominate. There appears to be an increase in the presence of extreme-\me\ BSSs as the cluster mass increases till $\sim$ 40000 \Mnom, as suggested by the panel (e). Irrespective of cluster mass, the low\me\ BSSs is the dominant population. The density parameter (panel (f)) shows no correlation with the maximum possible \me.

\begin{figure}
    \centering
    \includegraphics[width=0.48\textwidth]{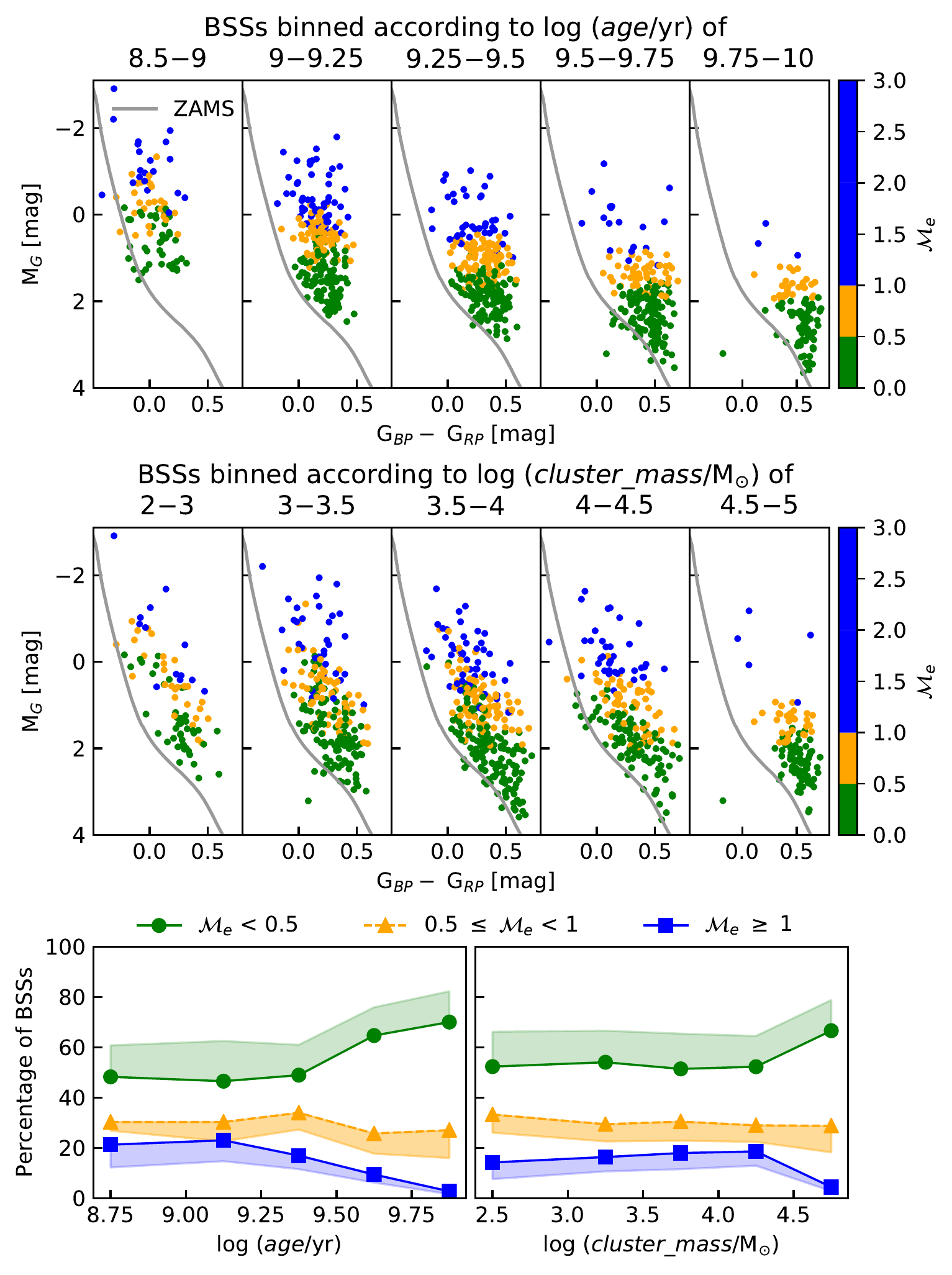}
    \caption{\textit{Top panels}: CMDs of BSSs across various age bins. \textit{Middle panels:} CMDs of BSSs across various cluster mass bins. The ZAMS are shown in the CMDs for reference. The BSSs are coloured according to \me. \textit{Bottom panels:} The variation of percentage of stars in the three BSS classes for the above age and cluster mass bins. The shaded regions indicate the possible percentages if pBSs were included in the tally.}
    \label{fig:age_wise}
\end{figure}

\subsection{BSSs properties in age and mass binned clusters} \label{sec:agewise}
For a more in-depth and quantitative analysis across cluster ages and mass, we divided the BSSs in age and mass bins as mentioned in Table~\ref{tab:binwise}.
The top panels in Fig.~\ref{fig:age_wise} show the CMDs of BSSs binned according to the cluster age. The middle panels show the CMDs of BSSs binned according to the cluster mass. All the BSSs are coloured according to their fractional mass excess and divided into three classes, as mentioned above. As seen in Fig.~\ref{fig:me_dist}, the most populous age range and mass range are 1--3 Gyr and 1000--30000 \Mnom\ respectively. In the upper panels, we see that the brightest BSSs are in the younger clusters, as expected, and becomes progressively fainter in older clusters. When we bin the clusters by mass, the BSSs have a similar range of brightness across the group, except for the most massive bin. There is tentative evidence for an increase in the low-\me\ BSSs as a function of age, whereas no such trend is visually noticed with respect to mass. 

In order to quantitatively assess the presence of the above trend, the percentage of binned BSSs in the three classes for the age/mass groups are tabulated in Table~\ref{tab:binwise} and shown in the bottom panels of Fig.~\ref{fig:age_wise}. The panels also include the possible change in each class if pBSs were included in the calculations. From the bottom panels, it becomes apparent that the fraction of extreme-\me\ BSSs decreases with age beyond 1 Gyr and has a maximum fraction near 1 Gyr age. On the other hand, the fraction of low-\me\ BSSs increases steadily with the cluster age, beyond 1-2 Gyr. The high-\me\ fraction follows the trend of the extreme-\me\ BSSs, as a function of age, though with a lesser degree of reduction. The trends seen here might suggest that the high and extreme-\me\ BSSs might belong to similar formation pathways and are similarly affected by cluster age. In contrast, the low-\me\ BSSs, which are likely to have a different formation pathway, are affected oppositely with age.
The cluster mass seems to have less impact on the fractional mass excess of the BSSs present. The low and extreme-\me\ BSSs shows a constant fraction in most of the bins, with a deviation in the most massive bin. The high-\me\ BSSs do not show any trend at all. 

\subsection{BSS segregation} \label{sec:segregation}

\begin{figure}
    \centering
    \includegraphics[width=0.48\textwidth]{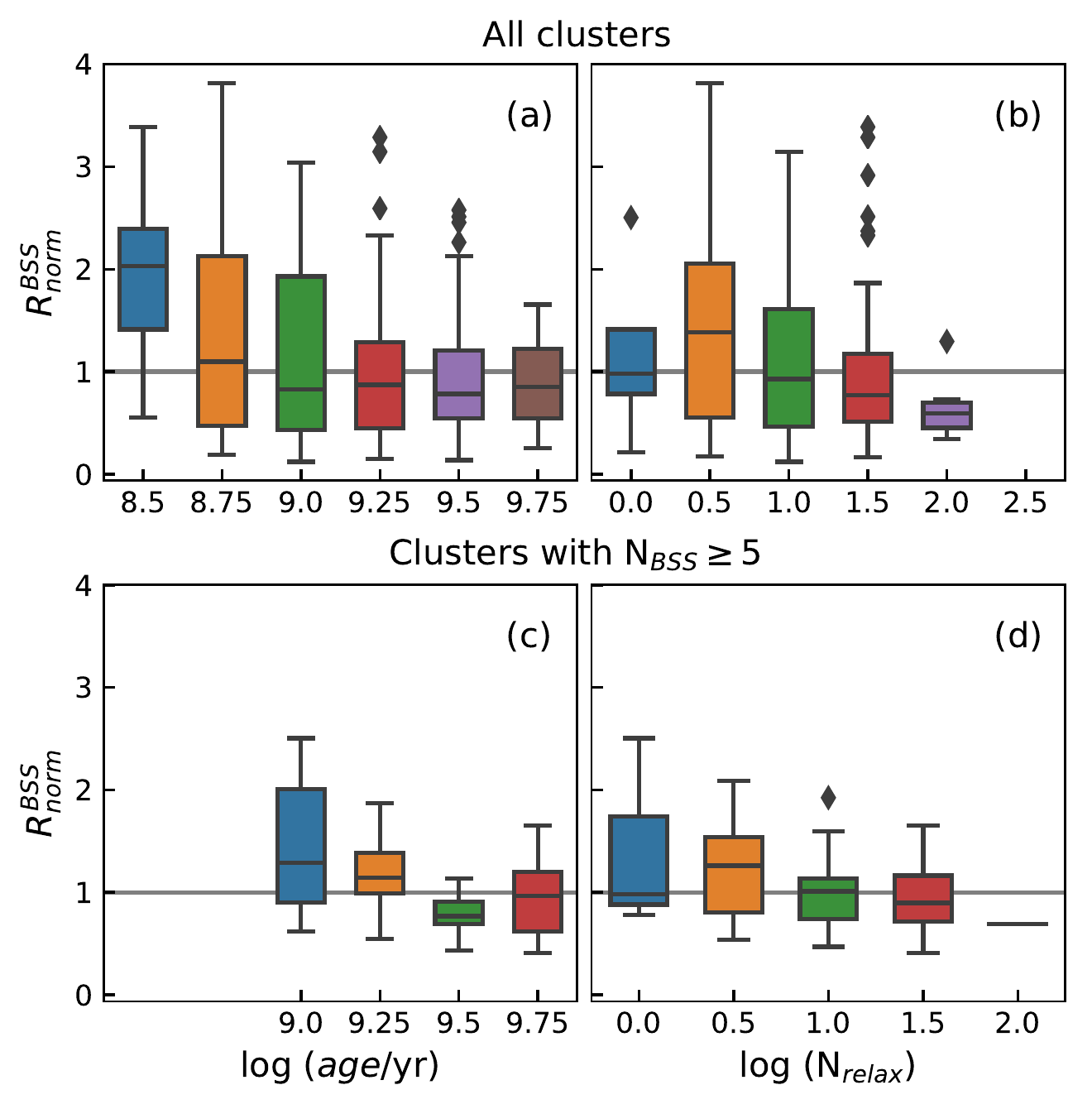}
    \caption{Box plots of the relationship of normalised effective radii of BSSs with the cluster age and dynamic relaxation time. The upper panels show the relationship for all clusters, while the lower panels only show for clusters with 5 or more BSSs.}
    \label{fig:segregation}
\end{figure}

BSSs are the most massive stars in a cluster; hence they are known/expected to be segregated towards the centre. To analyse the radial segregation, we calculated the effective radii of the cluster and BSSs (see Appendix \ref{sec:eff_radius} for definitions). 
Fig.~\ref{fig:segregation} (a) shows the variation of normalised BSS radii ($R_{norm}^{BSS}$) with cluster age. If the normalised radius has a value less than 1, it means that there is radial segregation. On average, the BSSs in clusters older than 1 Gyr are segregated, though there is quite a large range in the values of the $R_{norm}^{BSS}$ for clusters of all ages. Note that there can be large errors in $R_{norm}^{BSS}$ due to the low number of BSSs in a cluster. The trend of decreasing $R_{norm}^{BSS}$ with age is more prominent in clusters with at least 5 BSSs (Fig.~\ref{fig:segregation} (c)).

Fig.~\ref{fig:segregation} (b) \& (d) both show the variation of $R_{norm}^{BSS}$ as a function of the dynamical relaxation. The increasing segregation in clusters with larger $N_{relax}$ is more clear in these plots when compared to Fig.~\ref{fig:segregation} (a) and (c). Unfortunately, there are only 3 BSSs per cluster on average. Hence, the individual effective radii are not statistically significant. The global results from Fig.~\ref{fig:segregation} (d) demonstrate that the segregation of BSSs, though a function of cluster age, is related to the dynamical evolution of the cluster. 

We also compared the radial distance of individual BSS with its \me. On average, BSSs with \me\ $<$ 1.5 are segregated. However, contrary to expectations, BSSs with \me\ $>$ 1.5 are not segregated. These are massive BSSs and therefore are expected to be segregated. This would then indicate that these stars are not forming from the already segregated stars, but from the non-segregated systems. It is possible that these are short-lived and hence may evolve before getting radially segregated. The (g) panel of Fig.~\ref{fig:me_dist} shows the relation between \me\ and log(N$_{relax})$. The figure shows that the extreme-\me\ BSSs decreases with the dynamical relaxation of the clusters.
The decline of extreme-\me\ BSSs in relaxed clusters supports that they have not had enough time to segregate. The above points could provide some constraints/pointers to the formation pathways of extreme-\me\ BSSs.

\subsection{Probable BSS and data limitations}
The probable BSS (pBS) classification was devised to select all possible BSSs which can be very close to the MSTO while not contaminating the larger BS sample with MS-tip or MSTO stars. This group has both genuine BSSs and some MS stars. Therefore, one needs to be cautious while using the pBSs data in our catalogue.

Here we point out the potential issues which lead to incompleteness in this BS catalogue: These are either issues with the catalogue itself, such as calibration issues in G $<$ 12 mag, faintness cutoff of membership catalogue at G = 18 mag, missing proper motion and hence missing cluster members, or reduction of the cluster sample due to missing clusters near Galactic centre. A more detailed discussion on the \textit{Gaia} DR2 data can be found in \S\ 2.6 of \citet{Rain_2021arXiv210306004R}.

\subsection{Contamination in the BSS regime} \label{sec:contamination}
The BSS region has been known to overlap with the expected location of HB \citep{Bond1971PASP...83..638B, Jadhav2021arXiv210213375J}. We can estimate the number of HB stars in these clusters assuming the ratio of BSSs and HB stars in OCs is similar to globular clusters \citep{Leigh2009MNRAS.399L.179L}:
\begin{equation}
    \textrm{log}\frac{N_{HB}}{N_{BSS}} \approx 0.44\ \textrm{log} \frac{N_{TO+1}}{1000} - 0.36
\end{equation}
where $N_{TO+1}$ are the stars within 0 to 1 mag fainter than MSTO. HB stars have $M_G\sim1$ mag (from BASTI horizontal branch models; \citealt{Hidalgo2018ApJ...856..125H}). Among our sample, 157 clusters with BSS older have an MSTO fainter than 1 mag. Based on the above equation, there can be $\sim$49 HB stars in these 157 clusters. However, only blue HB (BHB) stars can be confused with BSSs. Assuming $B_{BHB}/B_{HB}$ of 0.25 \citep{Arimoto1981Ap&SS..76...73A}, there can be $\sim$12 BHB stars in our sample of 878 BSSs (1.4\%). As this is a very small fraction, the general results presented here are not affected.

There are two BSSs (one each in NGC 6791 and Haffner 5) that are quite bluer than the ZAMS (bottom left corner of Fig.~\ref{fig:BSS_CMDs} (c)). The star in Haffner 5 is very close to another bright star, while the star in NGC 6791 is likely a background B-type star \citep{Reed2012MNRAS.427.1245R}. These two stars do not impact the overall analysis presented here, but more caution is advised while studying individual BSSs in clusters.

The mass estimation assumed the BSSs are single stars. However, a large percentage of BSSs are known to be in binaries: 16/21 in NGC 188 \citep{Geller2008AJ....135.2264G,Mathieu2009Natur.462.1032M}; 11/14 in NGC 2682 \citep{Geller2015ApJ...808L..25G}; 6/12 in NGC 6791 \citep{Tofflemire2014AJ....148...61T}; 7/17 in NGC 6819 \citep{Milliman2014AJ....148...38M}; 4/13 in NGC 7789 \citep{Nine2020AJ....160..169N}. However, most of the binaries are single lined spectroscopic binaries (SB1). Among the 5 clusters, the fraction of double lined spectroscopic binaries (SB2s) is 4/77 (5.2\%). The shift in the CMD will be significant for SB2 systems, leading to overestimation of BSS mass. Assuming the SB2 percentage of BSSs in these 5 clusters is similar to the complete sample, 5.2\% (45 BSSs and 26 pBS) of the BSS candidates have overestimated mass.


\section{Discussion} \label{sec:discussion}
Out of 670 clusters in the age range 300 Myr to 10 Gyr, 228 OCs have bona fide BSSs, while 76 more clusters have probable BSSs. The number of BSSs roughly increases with the age and mass of the cluster. Here we discuss possible formation pathways suggested by this sample of BSSs and compare our findings with those in the literature.

\subsection{Reference mass for calculating \texorpdfstring{\me}{Me}}
\label{sec:reference_mass}
Understanding the relation between the mass of BSSs and the cluster turnoff is essential to put boundaries on the formation pathways. The most massive stars present in the cluster at the formation time of BSS tell about the possible pathways. However, the best estimation we can have regarding the most massive stars in the cluster is the cluster MSTO. This includes the assumption that the BSSs are not too old and the MSTO has not changed significantly during the lifetime of the BSSs. 
However, there are two ways to calculate the MSTO mass: (i) We can use the isochrone corresponding to the cluster age and get the MSTO mass (ii) We can use ZAMS and estimate the mass corresponding to the absolute G magnitude of the MSTO. The MSTO mass from the second method is larger than the first method.
Using the first method will lead to overestimating the \me\ values. Hence, to be conservative, we have used the second approach. 

We would like to note another possibility, that is, using the ZAMS mass corresponding to the MS-tip. In general, for clusters older than 1 Gyr, the mass as per the isochrone of cluster age at MSTO and MS-tip will be very similar. However, the mass from ZAMS will be much higher for MS-tip when compared to MSTO. Therefore, using ZAMS mass at MS-tip will be the most conservative approach, whereas the isochrone mass at the MSTO will be the least conservative approach. We also note that BSSs that are bluer but with similar brightness as MSTO will have negative excess mass if we consider the ZAMS mass at MS-tip as a reference. Therefore, we decided to use the moderately conservative ZAMS mass at MSTO as the reference. This decision does not impact the results much, as we do not attempt to pick each BSS and identify its formation pathway individually. On the other hand, we only attempt some statistics and trends concerning the fractional mass excess, \me.

\subsection{BSS formation pathways}
As \me\ is a proxy to the MT efficiency, we can roughly divide the BSSs into mergers and MT products. 
The low-\me\ BSSs encompass less-efficient MT binaries and merger products of low mass secondary. If we consider the typical mass-ratios of binaries to be $\sim0.7$, the mergers systems will contribute only a small portion in low-\me\ BSSs. High-\me\ BSSs are likely results of conservative MT and hence can be considered as merger dominated. The extreme-\me\ BSS have more mass than two MSTO stars, and therefore, they are either formed earlier when the MSTO mass was larger, or they are a product of more than 2 MSTO stars. The first scenario is less likely due to a very slow shift in $M_{TO}$ and the evolution of such BSSs away from the MS, particularly for M$_{TO}$ $<$ 2.0 clusters where these BSSs are identified. In summary, the majority of low-\me\ BSSs are likely to be $\sim$ MT products, whereas the majority of high-\me\ BSS are merger products, and extreme-\me\ BSS $\sim$ are likely to be multiple merger/MT products. We also found that the fraction of high and extreme-\me\ BSSs decreases with age. It is, therefore, possible that this is indicative of a reduction in the BSS formation through merger processes in clusters older than 1--2 Gyr. Moreover, the rising low-\me\ fraction could suggest a steadily increasing occurrence of BSSs formation through MT in clusters older than 1--2 Gyr. 

The extreme-\me\ BSSs apparently require more than twice the TO mass to be formed. They can be a result of unresolved binaries, gas accretion, multiple MT/merger events. As mentioned in \S~\ref{sec:contamination}, there is a 5.2\% chance that the BSSs are bright due to being part of unresolved binaries. This still leaves $\sim$130 extreme-\me\ BSSs which are likely single stars (or partner of a sub-luminous companion).
They have unlikely to be gained mass via gas accretion, as the clusters older than 5 Myr rarely have molecular gas \citep{Lada2003ARA&A..41...57L, Leisawitz1989ApJS...70..731L}.
\citet{Leigh2011MNRAS.410.2370L} have showed that 2+2, 2+3, 3+3 encounters are possible in the lifetimes of OCs. Moreover, multiple encounters can develop into mini-clusters of 5--7 stars, which can greatly increase the chance of collisions \citep{Geller2015ApJ...808L..25G}. Such encounters can also interrupt the MT for 20--40\% of the cases \citep{Leigh2016ApJ...818...21L}. Overall, the large percentage of extreme-\me\ BSSs indicates that multiple encounters are quite abundant in OCs.

\subsection{Comparison with literature}
The maximum number of BSSs have a power low dependence on cluster mass (N$_{BSS,max}$ $\alpha$ M$_{clus}^{0.6}$). Similar mass dependence was found in globular clusters by \citet{Knigge_2009Natur.457..288K} (N$_{BSS}$ $\alpha$ M$_{core}^{0.38\pm0.04}$), where they show a correlation between the number of BSSs and the core mass. The authors concluded that most BSSs, even those found in cluster cores of the globular clusters, come from binary systems. Similar to this, we found a power low fit to clusters with at least 10 BSSs (Fig.~\ref{fig:cl_params}). Also, the relation between the number of BSSs and the mass of the cluster is found to have a very similar value for the exponent (N$_{BSS}$ $\alpha$ M$_{clus}^{0.37 \pm 0.10}$). This would also suggest a binary origin for BSSs in open clusters, which is indeed true, as we find about 84\% of BSSs to be of binary origin. Therefore, the BSS formation and evolution is quite similar in open and globular clusters, both dominated by binaries.

\citet{Ahumada_2007A&A...463..789A} produced the most comprehensive list of BSSs before the arrival of \textit{Gaia}. Their list of 1887 candidate BSSs in 470 clusters can be considered an upper limit on the number of BSSs. Higher precision astrometry/membership with \textit{Gaia} will remove field contamination from their list. They had identified 148 (8\%) stars in 26 old clusters as high-mass BSSs. The fraction of extreme-\me\ BSSs (with \me\ $>$ 1) is 10--16\% from our estimates. This difference is likely due to the better classification using \textit{Gaia} and the differences in the cluster sample.

\citet{Leiner_2021ApJ...908..229L} analysed the BS population in 16 nearby old OCs. They defined $\delta M = M_{BSS}-M_{MSTO}$, where the stellar masses are calculated by comparison with isochrones. They analysed 16 clusters in the age range 1--10 Gyr; in comparison, this study used 165 clusters in the same age range. \citet{Leiner_2021ApJ...908..229L} stated that stars above $\delta M$ = 1 \Mnom\ are rare. Our estimates show that 12--20\% of the BSSs in the same cluster sample are extreme-\me\ BSSs. This discrepancy is the result of the definition of $M_{TO}$. We could replicate their Fig. 3 using the ZAMS mass of the MS-tip as the $M_{TO}$. As the MS-tip stars are roughly 1.2 mag brighter than the MSTO (for the old clusters), $M_{TO}$ used by \citet{Leiner_2021ApJ...908..229L} is significantly larger than our $M_{TO}$, resulting in a conservative estimate of the excess mass of BSSs. As mentioned before, their method produces negative excess mass for fainter (than the MS-tip) and bluer BSSs, making it difficult to classify them as BSSs, and understand their formation pathways.

\citet{Rain_2021arXiv210306004R} recently produced a BSS catalogue using the parent membership data from \citet{Cantat2020}, although for clusters of all ages. They found 899 BSS candidates in 408 OCs. They also found that BSSs are largely absent in clusters younger than 500 Myr. The difference between their catalogue and this work can come from the adopted age criteria, selection method and different membership probability cutoffs used in the two studies (50\% in \citealt{Rain_2021arXiv210306004R} and 70\% in this work).

We also note that though we have estimated the number of pBS stars, we do not use them in the analysis to find correlations with cluster properties. This is because these are only BSS candidates. The pBSs generally have low excess mass and are likely to be the product of MT. If they are indeed BSSs, then the fraction of BSSs formed via binary interactions will increase to 90\% and reducing the fraction requiring multiple mergers/MT. 

We plan to carry out detailed analyses of BSSs in the catalogue to obtain their stellar properties in the future. Furthermore, interesting clusters and BSSs identified in this study will be followed with ultraviolet (UVIT on AstroSat) and spectroscopic observations (3.6 m Devasthal Optical Telescope).

\section{Summary}  \label{sec:Conclusions}
This study aimed to explore the properties of BSSs as a function of cluster parameters and identify potential clusters for further study. The data of 868 BSSs in 228 clusters in the 0.3--10 Gyr age and 10$^2$--10$^4$ \Mnom\ mass range are used to explore their properties. The conclusions we derive from this study are the following:
\begin{enumerate}

\item The number of BSSs found in a cluster is dependent on the cluster age. We derive the average BSS/cluster for different age ranges, which shows an increasing trend with age (1.6 BSS/cluster (log(\textit{age}) 8.5--9.0); 3.4 (9.0--9.5); 7.9 (9.5--9.75) and 17.8 (9.75--10.0).

\item The maximum number of BSSs found in a cluster is related to the cluster mass by a power-law (N$_{BSS,max}$ $\alpha$ M$_{clus}^{0.6}$). The clusters with at least 10 BSSs show a power low relation (N$_{BSS}$ $\alpha$ M$_{clus}^{0.37 \pm 0.10}$) similar to globular clusters, indicating binary dominated BSS formation in both type of clusters.

\item The number of BSS found in a cluster is not dependent on the density, radius, distance, Galactocentric distance or number of stars near the turnoff. The BSS fraction has no correlation with age, but has positive correlation with respect to the cluster relaxation.

\item We introduced the term fractional mass excess for a BSS (\me\ $\sim$ BSS mass normalised to MSTO) to differentiate various formation pathways. We divided the BSSs into 3 groups: low-\me\ ($<$ 0.5), high-\me\ ($1.0 < \mathcal{M}_e < 0.5$) and extreme-\me\ ($>$ 1.0). We suggest that the low-\me\ group is likely to be formed via MT, high-\me\ to be formed via binary mergers and extreme-\me\ to be through multiple merger/MT. 
 
\item Majority of BSSs are formed by interactions in binaries, either by MT (as suggested by low-\me\ = 54\%) or merger in binaries (high-\me\ = 30\%). In addition, a not-so-big but significant fraction of BSSs is likely to be formed through multiple merger/MT, as suggested by 16\% of extreme-\me\ BSSs. 

\item The percentage of high and extreme-\me\ BSSs show a similar decreasing trend with age beyond 1--2 Gyr, possibly indicating a reduction in the BSS formation through merger processes in older clusters. The rising low-\me\ fraction beyond 1--2 Gyr could suggest a steadily increasing occurrence of BSS formation through MT beyond 1--2 Gyr.

\item The radial segregation of the BSSs is not found to be a direct function of age, but rather a direct function of the dynamical relaxation of the cluster. On average, BSSs with \me\ $<$ 1.5 are found to be segregated. Contrary to expectations, those with larger \me\ are comparatively less segregated, suggesting that they are not formed from the already segregated stars.
\end{enumerate}

\section*{Acknowledgements}
We thank the referee for constructive comments which helped in improving the manuscript.
This work has made use of data from the European Space Agency (ESA) mission {\it Gaia} (\url{https://www.cosmos.esa.int/gaia}), processed by the {\it Gaia} Data Processing and Analysis Consortium (DPAC, \url{https://www.cosmos.esa.int/web/gaia/dpac/consortium}). Funding for the DPAC has been provided by national institutions, in particular the institutions participating in the {\it Gaia} Multilateral Agreement.

\section*{Data Availability}
The Gaia DR2 data is available at \url{https://gea.esac.esa.int/archive/}. The list of clusters with BSSs (Table \ref{tab:cluster_list}) and list of all BSS candidates (Table \ref{tab:catalogue}) are available online as supplementary material and at CDS via anonymous ftp to cdsarc.u-strasbg.fr (130.79.128.5) or via \url{http://cdsarc.u-strasbg.fr/viz-bin/qcat?J/MNRAS}.

\bibliographystyle{mnras}
\bibliography{references} 

\begin{thebibliography}{}
\makeatletter
\relax
\def\mn@urlcharsother{\let\do\@makeother \do\$\do\&\do\#\do\^\do\_\do\%\do\~}
\def\mn@doi{\begingroup\mn@urlcharsother \@ifnextchar [ {\mn@doi@}
  {\mn@doi@[]}}
\def\mn@doi@[#1]#2{\def\@tempa{#1}\ifx\@tempa\@empty \href
  {http://dx.doi.org/#2} {doi:#2}\else \href {http://dx.doi.org/#2} {#1}\fi
  \endgroup}
\def\mn@eprint#1#2{\mn@eprint@#1:#2::\@nil}
\def\mn@eprint@arXiv#1{\href {http://arxiv.org/abs/#1} {{\tt arXiv:#1}}}
\def\mn@eprint@dblp#1{\href {http://dblp.uni-trier.de/rec/bibtex/#1.xml}
  {dblp:#1}}
\def\mn@eprint@#1:#2:#3:#4\@nil{\def\@tempa {#1}\def\@tempb {#2}\def\@tempc
  {#3}\ifx \@tempc \@empty \let \@tempc \@tempb \let \@tempb \@tempa \fi \ifx
  \@tempb \@empty \def\@tempb {arXiv}\fi \@ifundefined
  {mn@eprint@\@tempb}{\@tempb:\@tempc}{\expandafter \expandafter \csname
  mn@eprint@\@tempb\endcsname \expandafter{\@tempc}}}

\bibitem[\protect\citeauthoryear{{Ahumada} \& {Lapasset}}{{Ahumada} \&
  {Lapasset}}{2007}]{Ahumada_2007A&A...463..789A}
{Ahumada} J.~A.,  {Lapasset} E.,  2007, \mn@doi [\aap]
  {10.1051/0004-6361:20054590}, \href
  {https://ui.adsabs.harvard.edu/abs/2007A&A...463..789A} {463, 789}

\bibitem[\protect\citeauthoryear{{Arimoto} \& {Simoda}}{{Arimoto} \&
  {Simoda}}{1981}]{Arimoto1981Ap&SS..76...73A}
{Arimoto} N.,  {Simoda} M.,  1981, \mn@doi [\apss] {10.1007/BF00651243}, \href
  {https://ui.adsabs.harvard.edu/abs/1981Ap&SS..76...73A} {76, 73}

\bibitem[\protect\citeauthoryear{{Boffin}, {Hillen}, {Berger}, {Jorissen},
  {Blind}, {Le Bouquin}, {Miko{\l}ajewska}  \& {Lazareff}}{{Boffin}
  et~al.}{2014}]{Boffin2014A&A...564A...1B}
{Boffin} H.~M.~J.,  {Hillen} M.,  {Berger} J.~P.,  {Jorissen} A.,  {Blind} N.,
  {Le Bouquin} J.~B.,  {Miko{\l}ajewska} J.,   {Lazareff} B.,  2014, \mn@doi
  [\aap] {10.1051/0004-6361/201323194}, \href
  {https://ui.adsabs.harvard.edu/abs/2014A&A...564A...1B} {564, A1}

\bibitem[\protect\citeauthoryear{{Bond} \& {Perry}}{{Bond} \&
  {Perry}}{1971}]{Bond1971PASP...83..638B}
{Bond} H.~E.,  {Perry} C.~L.,  1971, \mn@doi [\pasp] {10.1086/129190}, \href
  {https://ui.adsabs.harvard.edu/abs/1971PASP...83..638B} {83, 638}

\bibitem[\protect\citeauthoryear{{Bressan}, {Marigo}, {Girardi}, {Salasnich},
  {Dal Cero}, {Rubele}  \& {Nanni}}{{Bressan}
  et~al.}{2012}]{Bressan2012MNRAS.427..127B}
{Bressan} A.,  {Marigo} P.,  {Girardi} L.,  {Salasnich} B.,  {Dal Cero} C.,
  {Rubele} S.,   {Nanni} A.,  2012, \mn@doi [\mnras]
  {10.1111/j.1365-2966.2012.21948.x}, \href
  {https://ui.adsabs.harvard.edu/abs/2012MNRAS.427..127B} {427, 127}

\bibitem[\protect\citeauthoryear{{Cantat-Gaudin} et~al.,}{{Cantat-Gaudin}
  et~al.}{2018}]{Cantat2018}
{Cantat-Gaudin} T.,  et~al., 2018, \mn@doi [\aap]
  {10.1051/0004-6361/201833476}, \href
  {https://ui.adsabs.harvard.edu/abs/2018A&A...618A..93C} {618, A93}

\bibitem[\protect\citeauthoryear{{Cantat-Gaudin} et~al.,}{{Cantat-Gaudin}
  et~al.}{2020}]{Cantat2020}
{Cantat-Gaudin} T.,  et~al., 2020, \mn@doi [\aap]
  {10.1051/0004-6361/202038192}, \href
  {https://ui.adsabs.harvard.edu/abs/2020A&A...640A...1C} {640, A1}

\bibitem[\protect\citeauthoryear{{Davies}, {Piotto}  \& {de Angeli}}{{Davies}
  et~al.}{2004}]{Davies_2004MNRAS.349..129D}
{Davies} M.~B.,  {Piotto} G.,   {de Angeli} F.,  2004, \mn@doi [\mnras]
  {10.1111/j.1365-2966.2004.07474.x}, \href
  {https://ui.adsabs.harvard.edu/abs/2004MNRAS.349..129D} {349, 129}

\bibitem[\protect\citeauthoryear{{Ferraro} et~al.,}{{Ferraro}
  et~al.}{2009}]{Ferraro_2009Natur.462.1028F}
{Ferraro} F.~R.,  et~al., 2009, \mn@doi [\nat] {10.1038/nature08607}, \href
  {https://ui.adsabs.harvard.edu/abs/2009Natur.462.1028F} {462, 1028}

\bibitem[\protect\citeauthoryear{{Gaia Collaboration} et~al.,}{{Gaia
  Collaboration} et~al.}{2016}]{Gaia2016A&A...595A...1G}
{Gaia Collaboration} et~al., 2016, \mn@doi [\aap]
  {10.1051/0004-6361/201629272}, \href
  {https://ui.adsabs.harvard.edu/abs/2016A&A...595A...1G} {595, A1}

\bibitem[\protect\citeauthoryear{{Gaia Collaboration} et~al.,}{{Gaia
  Collaboration} et~al.}{2018}]{Gaia2018A&A...616A...1G}
{Gaia Collaboration} et~al., 2018, \mn@doi [\aap]
  {10.1051/0004-6361/201833051}, \href
  {https://ui.adsabs.harvard.edu/abs/2018A&A...616A...1G} {616, A1}

\bibitem[\protect\citeauthoryear{{Geller} \& {Leigh}}{{Geller} \&
  {Leigh}}{2015}]{Geller2015ApJ...808L..25G}
{Geller} A.~M.,  {Leigh} N. W.~C.,  2015, \mn@doi [\apjl]
  {10.1088/2041-8205/808/1/L25}, \href
  {https://ui.adsabs.harvard.edu/abs/2015ApJ...808L..25G} {808, L25}

\bibitem[\protect\citeauthoryear{{Geller}, {Mathieu}, {Harris}  \&
  {McClure}}{{Geller} et~al.}{2008}]{Geller2008AJ....135.2264G}
{Geller} A.~M.,  {Mathieu} R.~D.,  {Harris} H.~C.,   {McClure} R.~D.,  2008,
  \mn@doi [\aj] {10.1088/0004-6256/135/6/2264}, \href
  {https://ui.adsabs.harvard.edu/abs/2008AJ....135.2264G} {135, 2264}

\bibitem[\protect\citeauthoryear{{Gosnell}, {Mathieu}, {Geller}, {Sills},
  {Leigh}  \& {Knigge}}{{Gosnell} et~al.}{2014}]{Gosnell2014ApJ...783L...8G}
{Gosnell} N.~M.,  {Mathieu} R.~D.,  {Geller} A.~M.,  {Sills} A.,  {Leigh} N.,
  {Knigge} C.,  2014, \mn@doi [\apjl] {10.1088/2041-8205/783/1/L8}, \href
  {https://ui.adsabs.harvard.edu/abs/2014ApJ...783L...8G} {783, L8}

\bibitem[\protect\citeauthoryear{{Gosnell}, {Leiner}, {Mathieu}, {Geller},
  {Knigge}, {Sills}  \& {Leigh}}{{Gosnell}
  et~al.}{2019}]{Gosnell2019ApJ...885...45G}
{Gosnell} N.~M.,  {Leiner} E.~M.,  {Mathieu} R.~D.,  {Geller} A.~M.,  {Knigge}
  C.,  {Sills} A.,   {Leigh} N. W.~C.,  2019, \mn@doi [\apj]
  {10.3847/1538-4357/ab4273}, \href
  {https://ui.adsabs.harvard.edu/abs/2019ApJ...885...45G} {885, 45}

\bibitem[\protect\citeauthoryear{{Hidalgo} et~al.,}{{Hidalgo}
  et~al.}{2018}]{Hidalgo2018ApJ...856..125H}
{Hidalgo} S.~L.,  et~al., 2018, \mn@doi [\apj] {10.3847/1538-4357/aab158},
  \href {https://ui.adsabs.harvard.edu/abs/2018ApJ...856..125H} {856, 125}

\bibitem[\protect\citeauthoryear{{Hills} \& {Day}}{{Hills} \&
  {Day}}{1976}]{Hills_1976ApL....17...87H}
{Hills} J.~G.,  {Day} C.~A.,  1976, \aplett, \href
  {https://ui.adsabs.harvard.edu/abs/1976ApL....17...87H} {17, 87}

\bibitem[\protect\citeauthoryear{{Jadhav}, {Pandey}, {Subramaniam}  \&
  {Sagar}}{{Jadhav} et~al.}{2021}]{Jadhav2021arXiv210213375J}
{Jadhav} V.~V.,  {Pandey} S.,  {Subramaniam} A.,   {Sagar} R.,  2021, \mn@doi
  [Journal of Astrophysics and Astronomy] {10.1007/s12036-021-09746-y}, 42

\bibitem[\protect\citeauthoryear{{Knigge}, {Leigh}  \& {Sills}}{{Knigge}
  et~al.}{2009}]{Knigge_2009Natur.457..288K}
{Knigge} C.,  {Leigh} N.,   {Sills} A.,  2009, \mn@doi [\nat]
  {10.1038/nature07635}, \href
  {https://ui.adsabs.harvard.edu/abs/2009Natur.457..288K} {457, 288}

\bibitem[\protect\citeauthoryear{{Lada} \& {Lada}}{{Lada} \&
  {Lada}}{2003}]{Lada2003ARA&A..41...57L}
{Lada} C.~J.,  {Lada} E.~A.,  2003, \mn@doi [\araa]
  {10.1146/annurev.astro.41.011802.094844}, \href
  {https://ui.adsabs.harvard.edu/abs/2003ARA&A..41...57L} {41, 57}

\bibitem[\protect\citeauthoryear{{Leigh} \& {Sills}}{{Leigh} \&
  {Sills}}{2011}]{Leigh2011MNRAS.410.2370L}
{Leigh} N.,  {Sills} A.,  2011, \mn@doi [\mnras]
  {10.1111/j.1365-2966.2010.17609.x}, \href
  {https://ui.adsabs.harvard.edu/abs/2011MNRAS.410.2370L} {410, 2370}

\bibitem[\protect\citeauthoryear{{Leigh}, {Sills}  \& {Knigge}}{{Leigh}
  et~al.}{2007}]{Leigh_2007ApJ...661..210L}
{Leigh} N.,  {Sills} A.,   {Knigge} C.,  2007, \mn@doi [\apj] {10.1086/514330},
  \href {https://ui.adsabs.harvard.edu/abs/2007ApJ...661..210L} {661, 210}

\bibitem[\protect\citeauthoryear{{Leigh}, {Sills}  \& {Knigge}}{{Leigh}
  et~al.}{2009}]{Leigh2009MNRAS.399L.179L}
{Leigh} N.,  {Sills} A.,   {Knigge} C.,  2009, \mn@doi [\mnras]
  {10.1111/j.1745-3933.2009.00749.x}, \href
  {https://ui.adsabs.harvard.edu/abs/2009MNRAS.399L.179L} {399, L179}

\bibitem[\protect\citeauthoryear{{Leigh}, {Knigge}, {Sills}, {Perets},
  {Sarajedini}  \& {Glebbeek}}{{Leigh}
  et~al.}{2013}]{Leigh_2013MNRAS.428..897L}
{Leigh} N.,  {Knigge} C.,  {Sills} A.,  {Perets} H.~B.,  {Sarajedini} A.,
  {Glebbeek} E.,  2013, \mn@doi [\mnras] {10.1093/mnras/sts085}, \href
  {https://ui.adsabs.harvard.edu/abs/2013MNRAS.428..897L} {428, 897}

\bibitem[\protect\citeauthoryear{{Leigh}, {Geller}  \& {Toonen}}{{Leigh}
  et~al.}{2016}]{Leigh2016ApJ...818...21L}
{Leigh} N. W.~C.,  {Geller} A.~M.,   {Toonen} S.,  2016, \mn@doi [\apj]
  {10.3847/0004-637X/818/1/21}, \href
  {https://ui.adsabs.harvard.edu/abs/2016ApJ...818...21L} {818, 21}

\bibitem[\protect\citeauthoryear{{Leiner} \& {Geller}}{{Leiner} \&
  {Geller}}{2021}]{Leiner_2021ApJ...908..229L}
{Leiner} E.~M.,  {Geller} A.,  2021, \mn@doi [\apj] {10.3847/1538-4357/abd7e9},
  \href {https://ui.adsabs.harvard.edu/abs/2021ApJ...908..229L} {908, 229}

\bibitem[\protect\citeauthoryear{{Leisawitz}, {Bash}  \&
  {Thaddeus}}{{Leisawitz} et~al.}{1989}]{Leisawitz1989ApJS...70..731L}
{Leisawitz} D.,  {Bash} F.~N.,   {Thaddeus} P.,  1989, \mn@doi [\apjs]
  {10.1086/191357}, \href
  {https://ui.adsabs.harvard.edu/abs/1989ApJS...70..731L} {70, 731}

\bibitem[\protect\citeauthoryear{{Leonard}}{{Leonard}}{1989}]{Leonard_1989AJ.....98..217L}
{Leonard} P. J.~T.,  1989, \mn@doi [\aj] {10.1086/115138}, \href
  {https://ui.adsabs.harvard.edu/abs/1989AJ.....98..217L} {98, 217}

\bibitem[\protect\citeauthoryear{{Mathieu} \& {Geller}}{{Mathieu} \&
  {Geller}}{2009}]{Mathieu2009Natur.462.1032M}
{Mathieu} R.~D.,  {Geller} A.~M.,  2009, \mn@doi [\nat] {10.1038/nature08568},
  \href {https://ui.adsabs.harvard.edu/abs/2009Natur.462.1032M} {462, 1032}

\bibitem[\protect\citeauthoryear{{Mathieu} \& {Geller}}{{Mathieu} \&
  {Geller}}{2015}]{Mathieu_2015ASSL..413...29M}
{Mathieu} R.~D.,  {Geller} A.~M.,  2015, {The Blue Stragglers of the Old Open
  Cluster NGC 188}.
p.~29, \mn@doi{10.1007/978-3-662-44434-4_3}

\bibitem[\protect\citeauthoryear{{McCrea}}{{McCrea}}{1964}]{McCrea_1964MNRAS.128..147M}
{McCrea} W.~H.,  1964, \mn@doi [\mnras] {10.1093/mnras/128.2.147}, \href
  {https://ui.adsabs.harvard.edu/abs/1964MNRAS.128..147M} {128, 147}

\bibitem[\protect\citeauthoryear{{Milliman}, {Mathieu}, {Geller}, {Gosnell},
  {Meibom}  \& {Platais}}{{Milliman}
  et~al.}{2014}]{Milliman2014AJ....148...38M}
{Milliman} K.~E.,  {Mathieu} R.~D.,  {Geller} A.~M.,  {Gosnell} N.~M.,
  {Meibom} S.,   {Platais} I.,  2014, \mn@doi [\aj]
  {10.1088/0004-6256/148/2/38}, \href
  {https://ui.adsabs.harvard.edu/abs/2014AJ....148...38M} {148, 38}

\bibitem[\protect\citeauthoryear{{Naoz} \& {Fabrycky}}{{Naoz} \&
  {Fabrycky}}{2014}]{Naoz_2014ApJ...793..137N}
{Naoz} S.,  {Fabrycky} D.~C.,  2014, \mn@doi [\apj]
  {10.1088/0004-637X/793/2/137}, \href
  {https://ui.adsabs.harvard.edu/abs/2014ApJ...793..137N} {793, 137}

\bibitem[\protect\citeauthoryear{{Nine}, {Milliman}, {Mathieu}, {Geller},
  {Leiner}, {Platais}  \& {Tofflemire}}{{Nine}
  et~al.}{2020}]{Nine2020AJ....160..169N}
{Nine} A.~C.,  {Milliman} K.~E.,  {Mathieu} R.~D.,  {Geller} A.~M.,  {Leiner}
  E.~M.,  {Platais} I.,   {Tofflemire} B.~M.,  2020, \mn@doi [\aj]
  {10.3847/1538-3881/abad3b}, \href
  {https://ui.adsabs.harvard.edu/abs/2020AJ....160..169N} {160, 169}

\bibitem[\protect\citeauthoryear{{Perets} \& {Fabrycky}}{{Perets} \&
  {Fabrycky}}{2009}]{Perets_2009ApJ...697.1048P}
{Perets} H.~B.,  {Fabrycky} D.~C.,  2009, \mn@doi [\apj]
  {10.1088/0004-637X/697/2/1048}, \href
  {https://ui.adsabs.harvard.edu/abs/2009ApJ...697.1048P} {697, 1048}

\bibitem[\protect\citeauthoryear{{Rain}, {Ahumada}  \& {Carraro}}{{Rain}
  et~al.}{2021}]{Rain_2021arXiv210306004R}
{Rain} M.~J.,  {Ahumada} J.~A.,   {Carraro} G.,  2021, \mn@doi [\aap]
  {10.1051/0004-6361/202040072}, \href
  {https://ui.adsabs.harvard.edu/abs/2021A&A...650A..67R} {650, A67}

\bibitem[\protect\citeauthoryear{{Reed}, {Baran}, {{\O}stensen}, {Telting}  \&
  {O'Toole}}{{Reed} et~al.}{2012}]{Reed2012MNRAS.427.1245R}
{Reed} M.~D.,  {Baran} A.,  {{\O}stensen} R.~H.,  {Telting} J.,   {O'Toole}
  S.~J.,  2012, \mn@doi [\mnras] {10.1111/j.1365-2966.2012.22054.x}, \href
  {https://ui.adsabs.harvard.edu/abs/2012MNRAS.427.1245R} {427, 1245}

\bibitem[\protect\citeauthoryear{{Sandage}}{{Sandage}}{1953}]{Sandage_1953AJ.....58...61S}
{Sandage} A.~R.,  1953, \mn@doi [\aj] {10.1086/106822}, \href
  {https://ui.adsabs.harvard.edu/abs/1953AJ.....58...61S} {58, 61}

\bibitem[\protect\citeauthoryear{{Shao} \& {Li}}{{Shao} \&
  {Li}}{2016}]{Shao2016ApJ...833..108S}
{Shao} Y.,  {Li} X.-D.,  2016, \mn@doi [\apj] {10.3847/1538-4357/833/1/108},
  \href {https://ui.adsabs.harvard.edu/abs/2016ApJ...833..108S} {833, 108}

\bibitem[\protect\citeauthoryear{{Sills} \& {Bailyn}}{{Sills} \&
  {Bailyn}}{1999}]{Sills_1999ApJ...513..428S}
{Sills} A.,  {Bailyn} C.~D.,  1999, \mn@doi [\apj] {10.1086/306840}, \href
  {https://ui.adsabs.harvard.edu/abs/1999ApJ...513..428S} {513, 428}

\bibitem[\protect\citeauthoryear{{Sindhu} et~al.,}{{Sindhu}
  et~al.}{2019}]{Sindhu2019ApJ...882...43S}
{Sindhu} N.,  et~al., 2019, \mn@doi [\apj] {10.3847/1538-4357/ab31a8}, \href
  {https://ui.adsabs.harvard.edu/abs/2019ApJ...882...43S} {882, 43}

\bibitem[\protect\citeauthoryear{{Spitzer} \& {Hart}}{{Spitzer} \&
  {Hart}}{1971}]{Spitzer1971ApJ...164..399S}
{Spitzer} Lyman J.,  {Hart} M.~H.,  1971, \mn@doi [\apj] {10.1086/150855},
  \href {https://ui.adsabs.harvard.edu/abs/1971ApJ...164..399S} {164, 399}

\bibitem[\protect\citeauthoryear{{Subramaniam}, {Sagar}  \&
  {Bhatt}}{{Subramaniam} et~al.}{1993}]{Subramaniam1993A&A...273..100S}
{Subramaniam} A.,  {Sagar} R.,   {Bhatt} H.~C.,  1993, \aap, \href
  {https://ui.adsabs.harvard.edu/abs/1993A&A...273..100S} {273, 100}

\bibitem[\protect\citeauthoryear{{Subramaniam} et~al.,}{{Subramaniam}
  et~al.}{2016}]{Subramaniam2016ApJ...833L..27S}
{Subramaniam} A.,  et~al., 2016, \mn@doi [\apjl] {10.3847/2041-8213/833/2/L27},
  \href {https://ui.adsabs.harvard.edu/abs/2016ApJ...833L..27S} {833, L27}

\bibitem[\protect\citeauthoryear{{Tofflemire}, {Gosnell}, {Mathieu}  \&
  {Platais}}{{Tofflemire} et~al.}{2014}]{Tofflemire2014AJ....148...61T}
{Tofflemire} B.~M.,  {Gosnell} N.~M.,  {Mathieu} R.~D.,   {Platais} I.,  2014,
  \mn@doi [\aj] {10.1088/0004-6256/148/4/61}, \href
  {https://ui.adsabs.harvard.edu/abs/2014AJ....148...61T} {148, 61}

\bibitem[\protect\citeauthoryear{{de Mink}, {Pols}  \& {Hilditch}}{{de Mink}
  et~al.}{2007}]{Mink2007A&A...467.1181D}
{de Mink} S.~E.,  {Pols} O.~R.,   {Hilditch} R.~W.,  2007, \mn@doi [\aap]
  {10.1051/0004-6361:20067007}, \href
  {https://ui.adsabs.harvard.edu/abs/2007A&A...467.1181D} {467, 1181}

\makeatother
\end{thebibliography}


\appendix

\section{Calculations of stellar and cluster parameters} \label{sec:appendix_A}

\begin{figure*}
    \centering
    \includegraphics[width=0.48\textwidth]{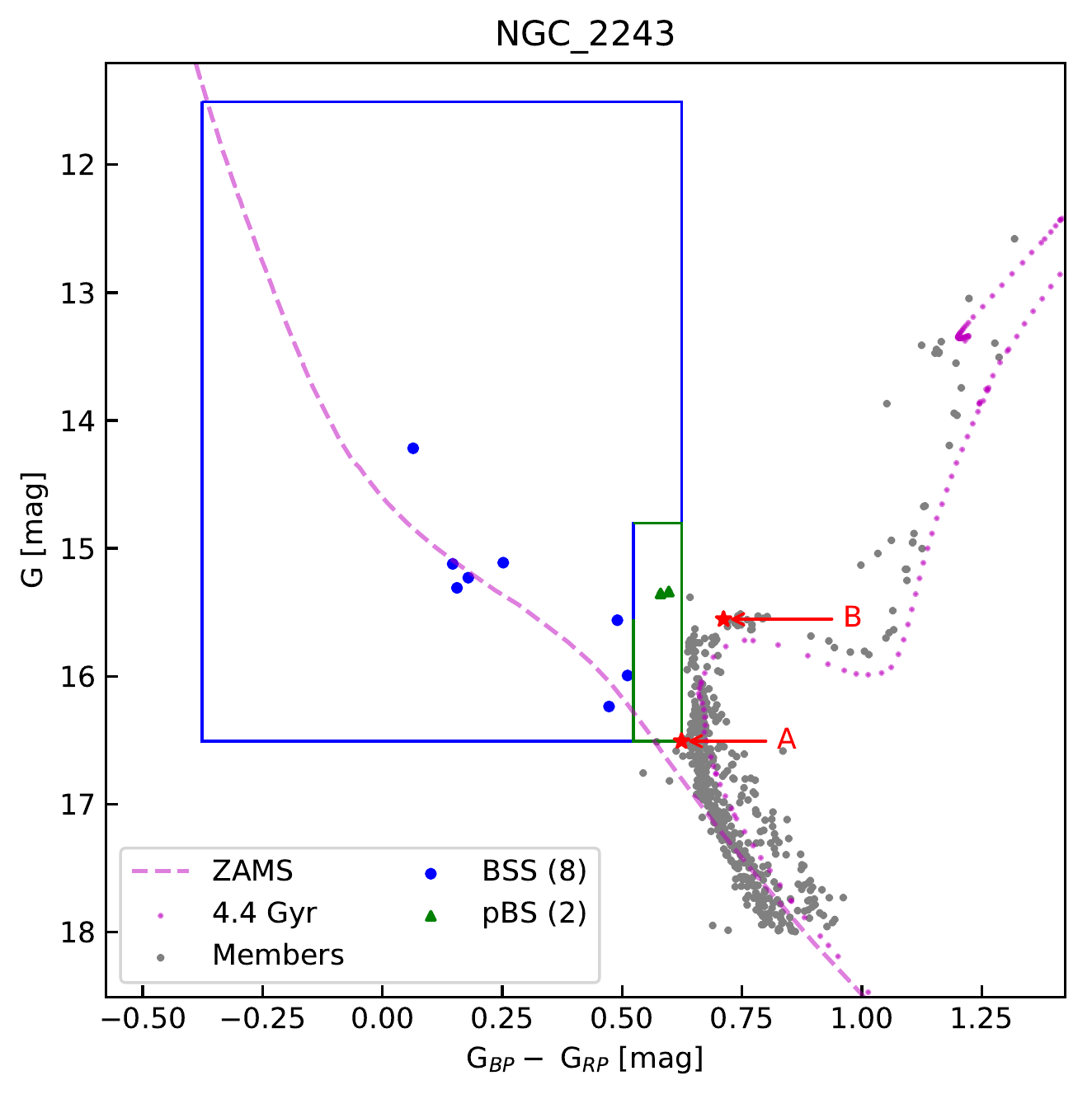}
    \includegraphics[width=0.48\textwidth]{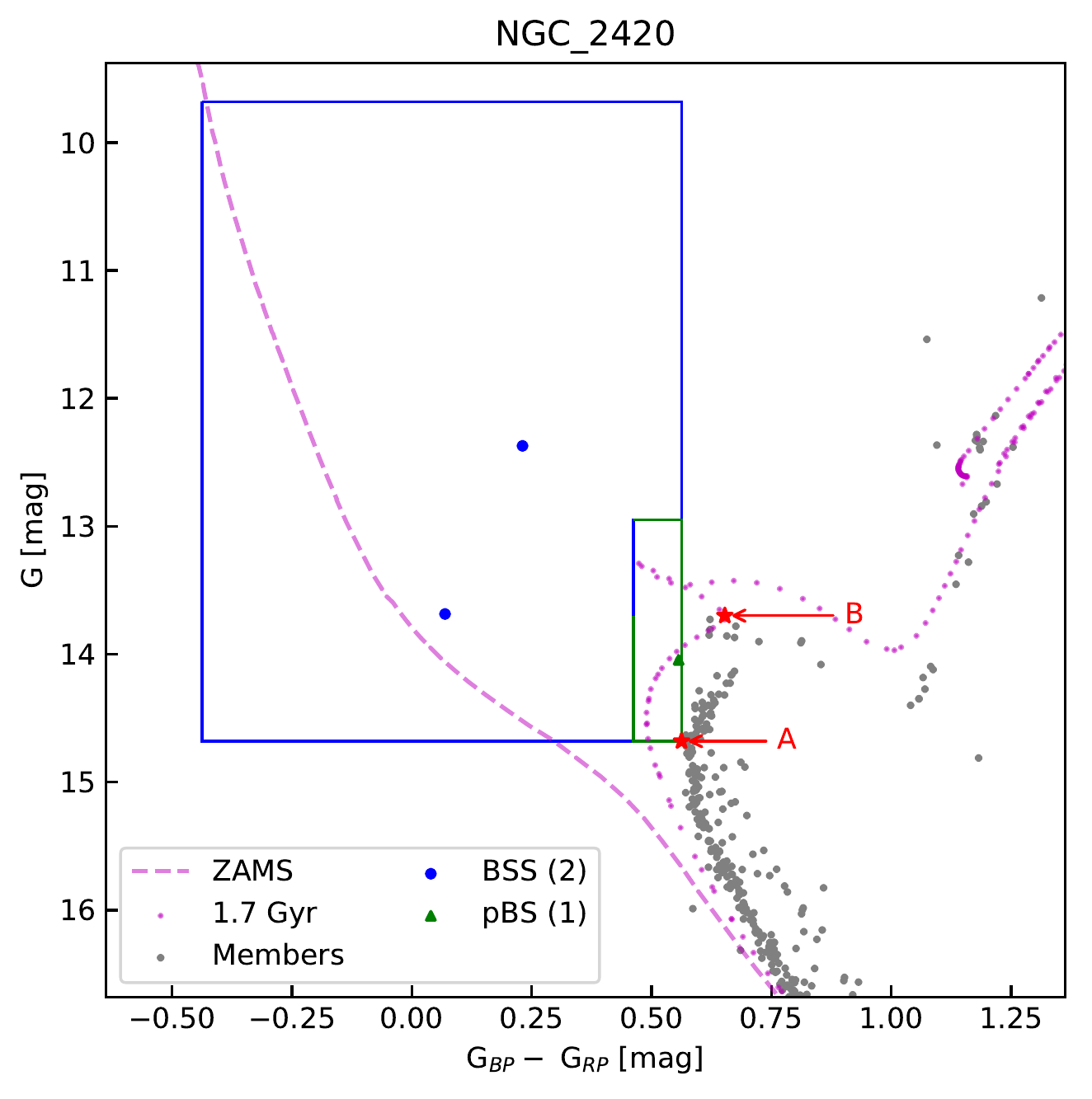}
    \includegraphics[width=0.48\textwidth]{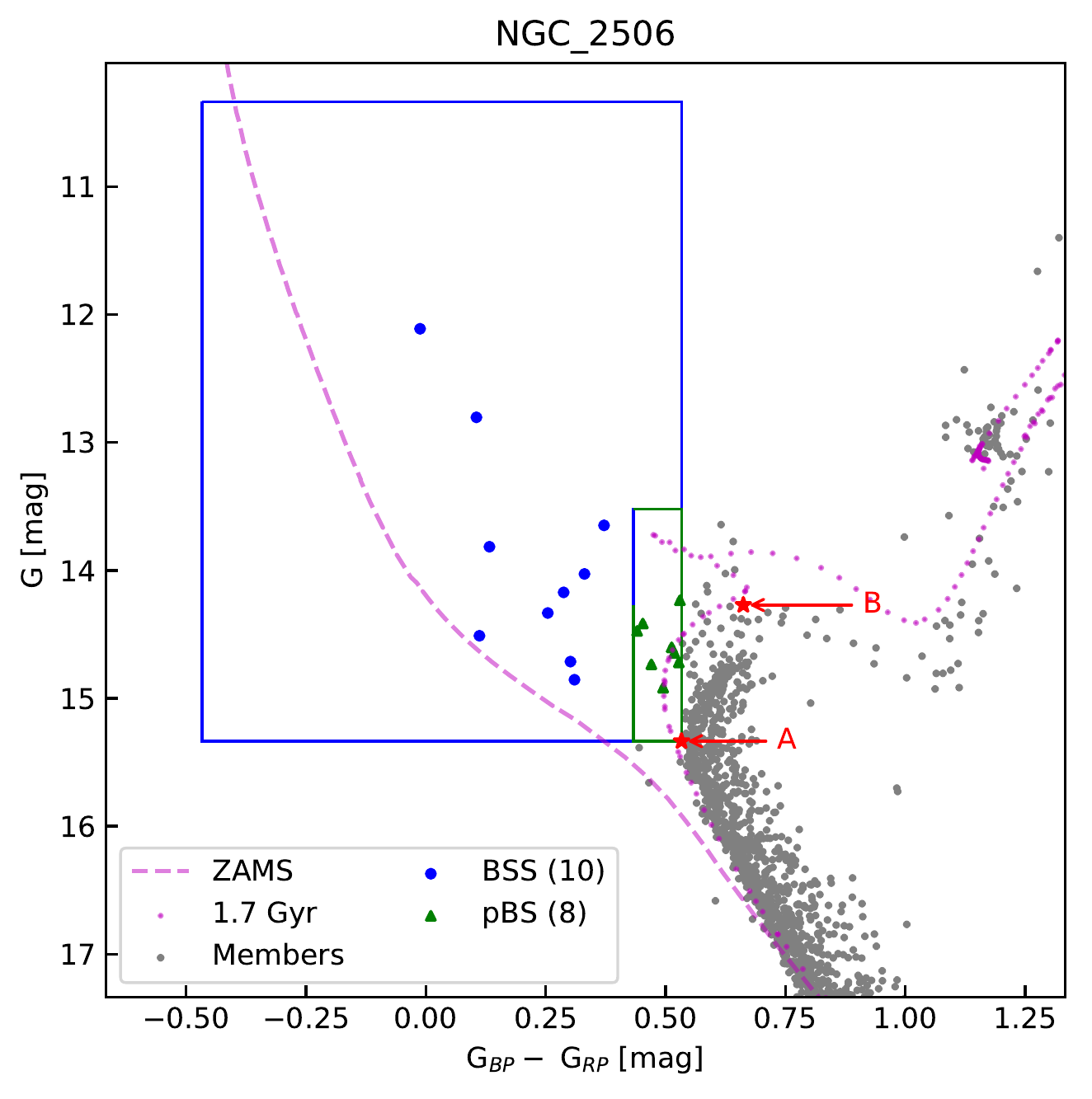}
    \includegraphics[width=0.48\textwidth]{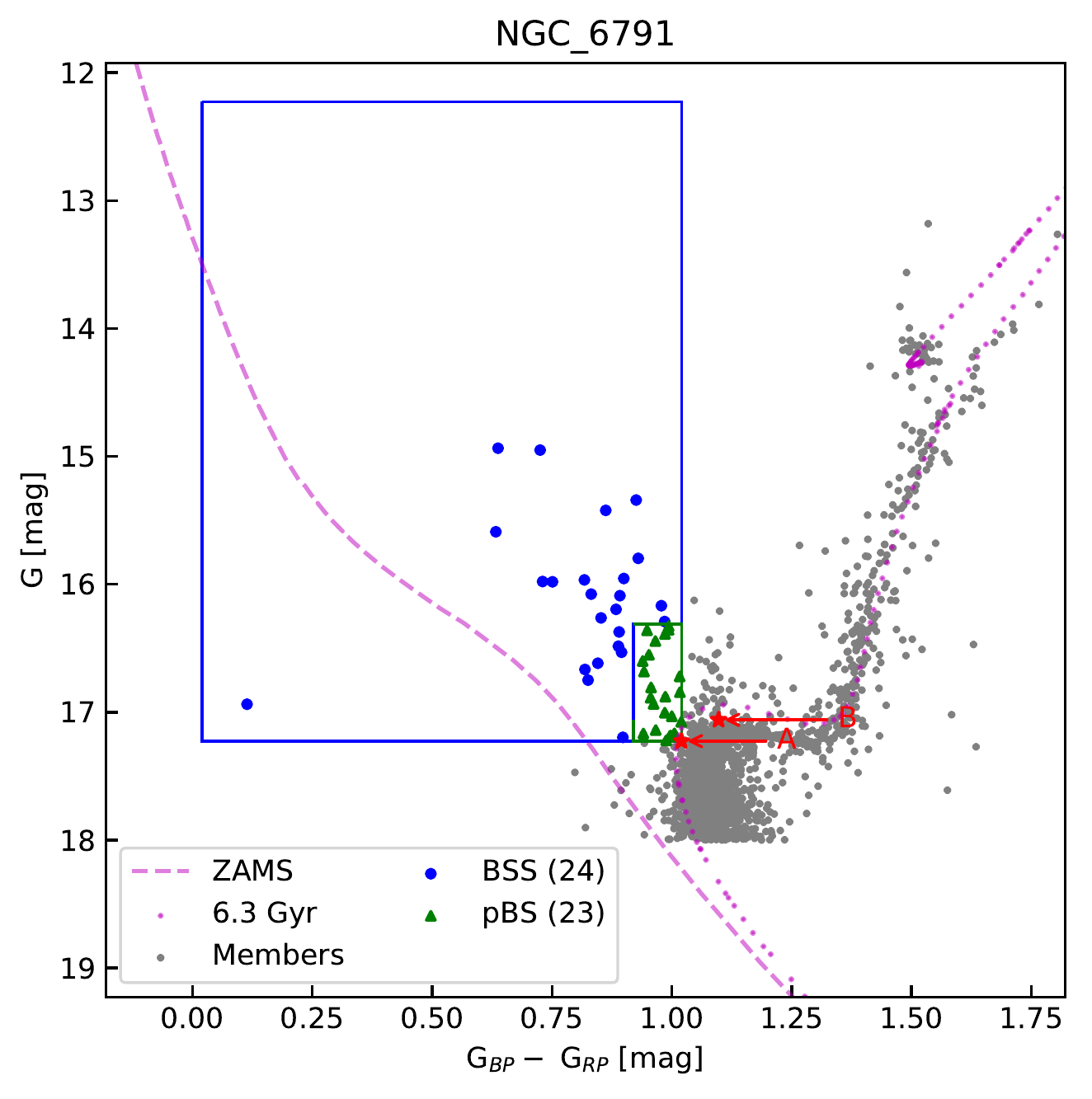}
    \caption{Example CMDs of NGC 2243, 2420, 2506 and 6791 showing selection of BSSs and pBSs. All elements are similar to Fig.~\ref{fig:demo_class}.}
    \label{fig:example_CMDs}
\end{figure*}

\begin{table*}
    \centering
    \begin{tabular}{ccccc ccccc cc}
    \toprule
Cluster	&	log($age$)	&	AV	&	DM	&	r	&	cl\_mass	&	density	&	$N_{relax}$	&	$N_{BSS}$	&	$N_{pBS}$	&	$R_{eff}^{cluster}$	&	$R_{eff}^{BSS}$	\\
	&		&		&		&	[pc]	&	[$M_{\odot}$]	&	[pc$^{-3}$]	&		&		&		&	[\arcmin]	&	[\arcmin]	\\ \hline
UBC\_199	&	9.06	&	0.93	&	10.46	&	2.9	&	300	&	4.37	&	16.13	&	0	&	0	&	10.77	&		\\
Skiff\_J0507+30.8	&	9.39	&	0.98	&	13.92	&	8.0	&	2626	&	1.81	&	3.70	&	2	&	2	&	4.48	&	9.54	\\
Czernik\_18	&	8.72	&	1.34	&	10.61	&	1.2	&	256	&	56.28	&	29.66	&	0	&	0	&	3.32	&		\\
COIN-Gaia\_11	&	8.9	&	1.25	&	9.13	&	3.9	&	304	&	1.81	&	7.05	&	1	&	0	&	20.58	&	26.72	\\
Berkeley\_69	&	8.9	&	1.61	&	12.66	&	2.3	&	1211	&	37.57	&	10.21	&	0	&	0	&	2.34	&		\\
\bottomrule
    \end{tabular}
    \caption{Example list of all clusters with log($age$) $>$ 8.5. Full table is available online.}
    \label{tab:cluster_list}
\end{table*}

\begin{table*}
    \centering
    \begin{tabular}{cccc cccc c}
    \toprule
Cluster	&	RA\_ICRS	&	DE\_ICRS	&	sourceID\_GaiaDR2	&	Gmag	&	BP-RP	&	$M_{BSS}$	&	\me	&	class	\\
	&	[deg]	&	[deg]	&		&		&		&	[$M_{\odot}$]	&		&		\\ \hline
Skiff\_J0507+30.8	&	76.96513	&	30.70442	&	156450304087516928	&	15.71	&	0.60	&	2.74	&	0.44	&	BSS	\\
Skiff\_J0507+30.8	&	76.67498	&	30.79202	&	157204805285086464	&	15.42	&	0.73	&	3.00	&	0.58	&	BSS	\\
COIN-Gaia\_11	&	68.60678	&	39.25254	&	178951225434386944	&	10.73	&	0.64	&	3.20	&	0.33	&	BSS	\\
Czernik\_21	&	81.65730	&	36.01864	&	184098241227893248	&	15.56	&	1.36	&	3.00	&	0.82	&	BSS	\\
Teutsch\_2	&	85.51016	&	39.12918	&	190160398586969984	&	14.83	&	0.72	&	3.40	&	0.62	&	BSS	\\

\bottomrule
    \end{tabular}
    \caption{Example list of all BSSs in our sample. A complete list of all BSSs and pBSs is available online.}
    \label{tab:catalogue}
\end{table*}

\subsection{Absolute magnitudes of the stars} \label{sec:absolute_mag}
Although the isochrones are not a perfect fit for the cluster, the distances and extinction given in \citet{Cantat2020} are accurate enough for statistical analysis. Hence, we calculate the absolute magnitudes using the cluster distance modulus (DM) and Av. Similarly, the absolute magnitudes of the MSTO point are also calculated.

\subsection{Mass of the stars} \label{sec:star_mass}
We defined a ZAMS using a combination of three isochrones. log($age$) = 8.00 for stars fainter than 2.5 G mag, log($age$) = 7.50 for stars within 2.5--1.5 G mag and log($age$) = 7.00 for stars within 1.5--4.0 G mag. These magnitudes cutoffs ensure a smooth transition in the CMD and G--Mass relation. The BSSs are assumed to be single stars, and their mass is estimated by comparing their absolute magnitudes to the ZAMS. The isochrones do not completely overlap with the cluster CMD, so there are errors associated with the mass estimation. We do not recommend this method for calculating the mass of an individual BSS. However, over the large sample, this method gives a rough estimate of the mass of the stars. Furthermore, the mass hierarchy within a cluster is unaffected by the fitting of the isochrone. 

\subsection{Mass and LF of the cluster} \label{sec:cl_mass}
The cluster mass is calculated by comparing the LF of the cluster with a model of the same age, distance and extinction and mass of 1 M$_{\odot}$. All clusters in the \citet{Cantat2020} catalogue are limited to 18 G mag, which gives the number of visible members ($LF_{18,\ cluster}$).

\begin{equation}
    cl\_mass = \frac{LF_{18,\ cluster}}{LF_{18,\ model}}\ [M_{\odot}]
\end{equation}
where, $LF_{18,\ model}$ is the LF of model cluster limited to apparent 18 G mag.

The total number of cluster members are derived using the complete model LF ($LF_{\infty,\ model}$) as,
\begin{equation}
    total\_stars = \frac{LF_{18,\ cluster} \times LF_{\infty,\ model}}{LF_{18,\ model}}
\end{equation}

The $total\_stars$ is dominated by very faint and small stars, which are not much important in the context of BSS evolution. Hence, we also calculated the number of stars near the turnoff using a similar technique. Finally, the $TO\_stars$ are defined as the number of MS stars within 0 to 3 mag of the MSTO. 
\begin{equation}
    TO\_stars = \frac{LF_{TO+3,\ cluster}}{LF_{\infty,\ model}} \times total\_stars
\end{equation}

We also calculated the cluster number density using the half member radius ($r_{50}$) of the cluster and the number of $TO\_stars$ as follows:
\begin{equation}
    density = \frac{TO\_stars}{4/3\ \pi r_{50}^3} [pc^{-3}]
\end{equation}

\subsection{Effective radius} \label{sec:eff_radius}
The effective radii of the cluster are calculated as follows: 

\begin{equation}
        R_{eff}^{cluster} = \frac{\Sigma r_{cluster}}{N_{cluster}}
\end{equation}
\noindent where $r_{sample}$ are the individual radii and $N_{sample}$ is the total number of stars in the sample. The normalised radius for individual stars is calculated as:
\begin{equation}
        r_{norm}^{sample} = r_{sample} \big/ R_{eff}^{cluster}
\end{equation}
\noindent And the normalised effective radius of the population (e.g. BSSs) is calculated as:
\begin{equation}
        R_{norm}^{sample} = \frac{\Sigma r_{sample}}{N_{sample}} \bigg/ R_{eff}^{cluster}
\end{equation}

\subsection{Dynamical relaxation time of the cluster} \label{sec:relaxation}
The segregation of BSSs has been linked to the number of relaxation periods passed during the cluster age. We calculated the dynamical relaxation time for the cluster as follows \citep{Spitzer1971ApJ...164..399S, Subramaniam1993A&A...273..100S}:
\begin{align}
        T_{relax} &= \frac{8.9 \times 10^5 (N_{cluster}\ r_{50}^3)^{0.5} }{\langle m \rangle^{0.5} \textrm{log}(0.4N_{cluster})} [yr] \\
        N_{relax} &= Age/T_{relax}
\end{align}
\noindent where $r_{50}$ is the radius containing half members (which is taken as a substitute for half mass radius) and $\langle m \rangle$ is the average mass of the stars in the cluster ($=cl\_mass/total\_stars$).


\bsp	
\label{lastpage}
\end{document}